\documentclass[12pt,english,draftcls, one column]{IEEEtran}
\usepackage[T1]{fontenc}
\usepackage[latin9]{inputenc}
\usepackage{float}
\usepackage{amsthm}
\usepackage{amsmath}
\usepackage{amssymb}
\usepackage{graphicx}

\makeatletter

\floatstyle{ruled}
\newfloat{algorithm}{tbp}{loa}
\providecommand{\algorithmname}{Algorithm}
\floatname{algorithm}{\protect\algorithmname}

\theoremstyle{plain}

\theoremstyle{plain}

\ifCLASSINFOpdf
\else
\fi
\usepackage{babel}

\makeatother

\usepackage{babel}
\providecommand{\propositionname}{Proposition}
\providecommand{\theoremname}{Theorem}

\begin{document}

\title{Multihop Backhaul Compression for the Uplink of Cloud Radio Access
Networks}

\author{Seok-Hwan Park, Osvaldo Simeone, Onur Sahin and Shlomo Shamai (Shitz)
\thanks{S.-H. Park and O. Simeone are with the Center for Wireless Communications
and Signal Processing Research (CWCSPR), ECE Department, New Jersey
Institute of Technology (NJIT), Newark, NJ 07102, USA (email: \{seok-hwan.park,
osvaldo.simeone\}@njit.edu).

O. Sahin is with InterDigital Inc., Melville, New York, 11747, USA
(email: Onur.Sahin@interdigital.com).

S. Shamai (Shitz) is with the Department of Electrical Engineering,
Technion, Haifa, 32000, Israel (email: sshlomo@ee.technion.ac.il).%
}}
\maketitle
\begin{abstract}
In cloud radio access networks (C-RANs), the baseband processing of
the radio units (RUs) is migrated to remote control units (CUs). This
is made possible by a network of backhaul links that connects RUs
and CUs and that carries compressed baseband signals. While prior
work has focused mostly on single-hop backhaul networks, this paper
investigates efficient backhaul compression strategies for the uplink
of C-RANs with a general multihop backhaul topology. A baseline multiplex-and-forward
(MF) scheme is first studied in which each RU forwards the bit streams
received from the connected RUs without any processing. It is observed
that this strategy may cause significant performance degradation in
the presence of a dense deployment of RUs with a well connected backhaul
network. To obviate this problem, a scheme is proposed in which each
RU decompresses the received bit streams and performs linear in-network
processing of the decompressed signals. For both the MF and the decompress-process-and-recompress
(DPR) backhaul schemes, the optimal design is addressed with the aim
of maximizing the sum-rate under the backhaul capacity constraints.
Recognizing the significant demands of the optimal solution of the
DPR scheme in terms of channel state information (CSI) at the RUs,
decentralized optimization algorithms are proposed under the assumption
of limited CSI at the RUs. Numerical results are provided to compare
the performance of the MF and DPR schemes, highlighting the potential
advantage of in-network processing and the impact of CSI limitations.\end{abstract}
\theoremstyle{theorem}
\newtheorem{theorem}{Theorem}
\theoremstyle{proposition}
\newtheorem{proposition}{Proposition}
\theoremstyle{lemma}
\newtheorem{lemma}{Lemma}
\theoremstyle{corollary}
\newtheorem{corollary}{Corollary}
\theoremstyle{definition}
\newtheorem{definition}{Definition}
\theoremstyle{remark}
\newtheorem{remark}{Remark}

\section{Introduction\label{sec:Introduction}}

\begin{figure}
\centering\includegraphics[width=14cm,height=9cm]{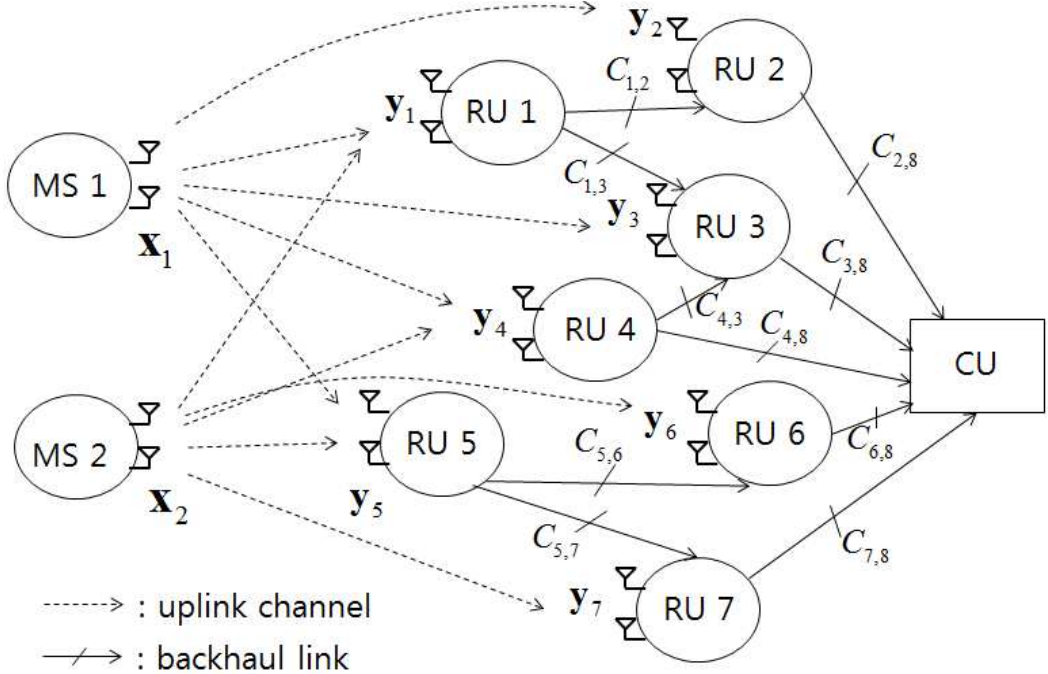}

\caption{{\footnotesize{\label{fig:system-model}Illustration of the uplink
of cloud radio access networks (C-RANs) with a multihop backhaul network
(RU: Radio Unit, CU: Control Unit).}}}
\end{figure}

Cloud radio access networks (C-RANs) \cite{Alcatel}\cite{China} prescribe
the separation of localized and distributed radio units (RUs) from
remote and centralized information processing nodes or control units
(CUs). The centralization of information processing afforded by C-RANs
potentially enables effective interference management at the geographical
scale covered by the distributed RUs. The main roadblock to the realization
of this potential hinges on the effective integration of the wireless
interface provided by the RUs with the backhaul network \cite{Biermann-et-al}\cite{IDT}.

With standard backhaul solutions based on the use of standard analog-to-digital
conversion techniques in the uplink and standard digital-to-analog
conversion techniques in the downlink \cite{CPRI}, backhaul capacity
limitations are known to impose a formidable bottleneck to the system
performance (see, e.g., \cite{IDT}). In order to alleviate the performance
bottleneck identified above, recent efforts by industry and academia
have targeted the design of more advanced \emph{backhaul compression}
schemes, which are based on \emph{point-to-point} vector compression
algorithms (see, e.g., \cite{Alcatel} and also \cite{Fettweis} for
experimental result). Following information-theoretic insights, \textit{multiterminal},
as opposed to point-to-point, backhaul compression techniques have
been studied in \cite{Sanderovich}-\cite{WeiYu} for the uplink and
in \cite{Park-et-al:TSP} for the downlink.

The research activity reviewed above assumes a single-hop, or \textit{star,}
backhaul topology in which each RU is directly connected to its managing
CUs via a backhaul link. In this work, instead, we study a more general
\textit{multihop} backhaul topology in which each RU may communicate
with the managing CU through a set of intermediate RUs as shown in
Fig. \ref{fig:system-model}. This backhaul topology is especially
relevant for heterogeneous small-cell networks in which RUs of various
sizes such as pico/femto or macro base stations are connected by a
mesh backhaul network \cite{Hur-et-al} (see also the standard \cite{CPRI}).

Reference \cite{Ni-et-al} provides a simulation-based study of the
performance of uplink C-RANs over multihop networks under the assumption
that each RU is able to evaluate the log-likelihood ratios  of the
transmitted bits of the connected mobile stations (MSs). In-network
processing of the log-likelihood ratios is proposed to enhance the
effectiveness of the use of the backhaul network. In this paper, we
instead focus on RUs that directly compress the received baseband
signal without performing any demodulation, following the standard
set-up for C-RAN (see, e.g., \cite{China}\cite{CPRI}). Reference
\cite{Goela-Gastpar} studies the related problem of optimizing linear
in-network processing operations in multihop network within the context
of \textit{estimation} (and not reliable digital communication). The
advantages of in-network processing were also investigated in \cite{Kumar-et-al}
for function computation in distributed sensor networks. We finally
point to related research activity on the performance of multihop
Gaussian relay networks with compress-and-forward strategies, single-antenna
nodes and fixed compression strategies, such as \cite{Avestimehr-et-al}-\cite{Sung-et-al}
(see also \cite{Krishna-et-al}\cite{Jiang-et-al}).

The paper organization and main contributions are as follows.
\begin{itemize}
\item In Sec. \ref{sec:System-Model}, we present the system model and describe
the general structure of backhaul routing strategies;
\item We investigate the \textit{Multiplex-and-Forward} (MF) scheme in Sec.
\ref{sec:Forward-without-Decompress}, whereby each RU forwards the
bit streams received from the connected RUs without any processing;
\item It is observed that this strategy may incur significant performance
degradation when the RUs have a sufficiently large number of incoming
backhaul links. In fact, in this case, the bit rate obtained by multiplexing
the signals received from the connected RUs is large and the backhaul
capacity constraints may impose a critical performance bottleneck;
\item We propose and investigate the \textit{Decompress-Process-and-Recompress}
(DPR) scheme that performs linear in-network processing of the compressed
baseband signals. The proposed DPR strategy performs the joint optimization
of the linear processing matrices and the compression strategies by
assuming that each RU has full channel state information (CSI);
\item Since the full CSI assumption at each RU may not be practical when
the number of RUs grows large, in Sec. \ref{sec:Decentralized-Optimization},
we propose decentralized DPR strategies whereby each RU computes its
linear processing and compression strategies using only local CSI;
\item We discuss an extension of the DPR scheme to the case in which there
are multiple CUs connected to each other on the backhaul network in
Sec. \ref{sub:Multiple-Control-Unit-Case};
\item Finally, in Sec. \ref{sec:Numerical-Results}, we provide extensive
numerical results to assess the performance of the considered schemes.
\end{itemize}
We conclude the paper in Sec. \ref{sec:Conclusion}.

\textit{Notation}: We adopt standard information-theoretic definitions
for the mutual information $I(X;Y)$ between the random variables
$X$ and $Y$, conditional mutual information $I(X;Y|Z)$ between
$X$ and $Y$ conditioned on random variable $Z$ \cite{ElGamal-Kim}.
All logarithms are in base two unless specified. The circularly symmetric
complex Gaussian distribution with mean $\mbox{\boldmath${\mu}$}$
and covariance matrix $\mathbf{R}$ is denoted by $\mathcal{CN}(\mbox{\boldmath${\mu}$},\bold{R})$.
The set of all $M\times N$ complex matrices is denoted by $\mathbb{C}^{M\times N}$,
and $\mathbb{E}[\cdot]$ represents the expectation operator. We use
the notation $\mathbf{X}\succeq\mathbf{0}$ to indicate that the matrix
$\mathbf{X}$ is positive semidefinite. The operation $(\cdot)^{\dagger}$
denotes Hermitian transpose of a matrix or vector, and notation $\mathbf{\Sigma}_{\mathbf{x}}$
is used for the correlation matrix of random vector $\mathbf{x}$,
i.e., $\mathbf{\Sigma}_{\mathbf{x}}=\mathbb{E}[\mathbf{x}\mathbf{x}^{\dagger}]$;
$\mathbf{\Sigma}_{\mathbf{x},\mathbf{y}}$ represents the cross-correlation
matrix $\mathbf{\Sigma}_{\mathbf{x},\mathbf{y}}=\mathbb{E}[\mathbf{x}\mathbf{y}^{\dagger}]$;
$\mathbf{\Sigma}_{\mathbf{x}|\mathbf{y}}$ is used for the conditional
correlation matrix, i.e., $\mathbf{\Sigma}_{\mathbf{x}|\mathbf{y}}=\mathbb{E}[\mathbf{x}\mathbf{x}^{\dagger}|\mathbf{y}]$,
and computed as $\mathbf{\Sigma}_{\mathbf{x}|\mathbf{y}}=\mathbf{\Sigma}_{\mathbf{x}}-\mathbf{\Sigma}_{\mathbf{x},\mathbf{y}}\mathbf{\Sigma}_{\mathbf{y}}^{-1}\mathbf{\Sigma}_{\mathbf{x},\mathbf{y}}^{\dagger}$.
Given a sequence of matrices $\mathbf{X}_{1},\ldots,\mathbf{X}_{m}$,
we define the notation $[\mathbf{X}_{1};\ldots;\mathbf{X}_{m}]=[\mathbf{X}_{1}^{\dagger},\ldots,\mathbf{X}_{m}^{\dagger}]^{\dagger}$
and the matrix $\mathbf{X}_{\mathcal{S}}$ for a subset $\mathcal{S}\subseteq\{1,\ldots,m\}$
as the matrix including, in ascending order, the matrices $\mathbf{X}_{i}$
with $i\in\mathcal{S}$.

\section{System Model\label{sec:System-Model}}

We consider the uplink of a C-RAN in which $N_{M}$ MSs transmit information
over a shared wireless medium to $N_{R}$ RUs as depicted in Fig.
\ref{fig:system-model}. The RUs are connected among themselves and
to the CUs that perform decoding of the MSs' information via a multihop
network of backhaul links. We define as $\mathcal{N}_{M}=\{1,\ldots,N_{M}\}$
and $\mathcal{N}_{R}=\{1,\ldots,N_{R}\}$ the sets of MSs and RUs,
respectively. MS $k$ and RU $i$ are equipped with $n_{M,k}$ and
$n_{R,i}$ antennas, respectively, for $k\in\mathcal{N}_{M}$ and
$i\in\mathcal{N}_{R}$. The total number of MSs' antennas is denoted
as $n_{M}=\sum_{k\in\mathcal{N}_{M}}n_{M,k}$. Fig. \ref{fig:system-model}
is an example with $N_{R}=7$ RUs, $N_{M}=2$ MSs, a single CU and
$n_{M,k}=n_{R,i}=2$ antennas at each terminal for $k\in\mathcal{N}_{M}$
and $i\in\mathcal{N}_{R}$.

\subsection{Channel Model}

Here we discuss the wireless uplink channel between MSs and RUs and
the multihop backhaul network connecting RUs and the CU. Specifically,
in most of the paper, we consider the case with a single CU, while
the more general scenario with multiple CUs is briefly treated in
Sec. \ref{sub:Multiple-Control-Unit-Case} (see Fig. \ref{fig:system-model-multi-CU}
for an illustration).

\textit{Uplink:} On the uplink channel, the signal $\mathbf{y}_{i}\in\mathbb{C}^{n_{R,i}\times1}$
received by RU $i$ at a given time is given by
\begin{equation}
\mathbf{y}_{i}=\mathbf{H}_{i}\mathbf{x}+\mathbf{z}_{i},\label{eq:observation}
\end{equation}
where $\mathbf{x}=[\mathbf{x}_{1};\mathbf{x}_{2};\ldots;\mathbf{x}_{N_{M}}]$
is the signal transmitted by all MSs with $\mathbf{x}_{k}\in\mathbb{C}^{n_{M,k}\times1}$
denoting the signal transmitted by MS $k$; $\mathbf{H}_{i}\in\mathbb{C}^{n_{R,i}\times n_{M}}$
is the flat-fading channel response matrix from all MSs toward RU
$i$; and $\mathbf{z}_{i}\in\mathbb{C}^{n_{R,i}\times1}$ is the additive
noise at RU $i$, which is distributed as $\mathbf{z}_{i}\sim\mathcal{CN}(\mathbf{0},\mathbf{I})$.
The signal $\mathbf{x}$ is distributed as $\mathbf{x}\sim\mathcal{CN}(\mathbf{0},\mathbf{\Sigma}_{\mathbf{x}})$
with covariance matrix $\mathbf{\Sigma}_{\mathbf{x}}=\mathrm{diag}(\mathbf{\Sigma}_{\mathbf{x}_{1}},\ldots,\mathbf{\Sigma}_{\mathbf{x}_{N_{M}}})$.
Note that the signals $\mathbf{x}_{k}$ are independent for $k\in\mathcal{N}_{M}$,
since the MSs are not able to cooperate. As a result, the signal $\mathbf{y}=[\mathbf{y}_{1};\mathbf{y}_{2};\ldots;\mathbf{y}_{N_{R}}]$
received by all RUs is distributed as $\mathbf{y}\sim\mathcal{CN}(\mathbf{0},\mathbf{\Sigma}_{\mathbf{y}})$
with $\mathbf{\Sigma}_{\mathbf{y}}=\mathbf{H}\mathbf{\Sigma}_{\mathbf{x}}\mathbf{H}^{\dagger}+\mathbf{I}$
and $\mathbf{H}=[\mathbf{H}_{1};\mathbf{H}_{2};\ldots;\mathbf{H}_{N_{R}}]$.

\textit{Backhaul network:} In order to model the backhaul multihop
network connecting the RUs and the CU, we define a capacitated directed
acyclic graph $\mathcal{G}=(\mathcal{V},\mathcal{E})$ (see, e.g.,
\cite{Koetter-Medard}). Accordingly, the set of vertex nodes of the
directed acyclic graph is $\mathcal{V}=\mathcal{N}_{R}\cup\{N_{R}+1\}$,
where the node $i$ represents the $i$th RU for $i\in\mathcal{N}_{R}$
and the last node $N_{R}+1$ stands for the CU. Also, the set $\mathcal{E}\subseteq\mathcal{V}\times\mathcal{V}$
contains the edges, where an edge $e=(i,j)$ represents the backhaul
link of capacity $C_{i,j}$ bits/s/Hz connecting node $i$ to node
$j$. The capacity $C_{i,j}$ is normalized by the bandwidth used
on the uplink wireless channel (as in, e.g., \cite{Zhang-et-al}).
Note that this enables the capacity $C_{i,j}$ to be equivalently
measured in bits per channel use of the uplink. The head and tail
of edge $e=(i,j)$ with respect to the direction $i\rightarrow j$
are denoted by $\mathrm{head}(e)=j$ and $\mathrm{tail}(e)=i$, respectively.

\begin{remark}In the given system model, all the RUs generally serve
the double purpose of radio receivers on the uplink and of intermediate
hops between ``upstream'' RUs and the CU on the backhaul network.
In practice, some nodes may not operate as radio receivers but only
as intermediate nodes in the backhaul network. This situation is captured
by the model by setting the channel matrix $\mathbf{H}_{i}$ in (\ref{eq:observation})
to have all-zero entries for all such nodes. In the following, we
hence refer to all nodes that belong to the backhaul network as RUs
with the understanding that some of them may only serve as relays.\end{remark}

\subsection{Backhaul Routing\label{sec:Routing}}

\begin{figure}
\centering\includegraphics[width=16cm,height=7cm]{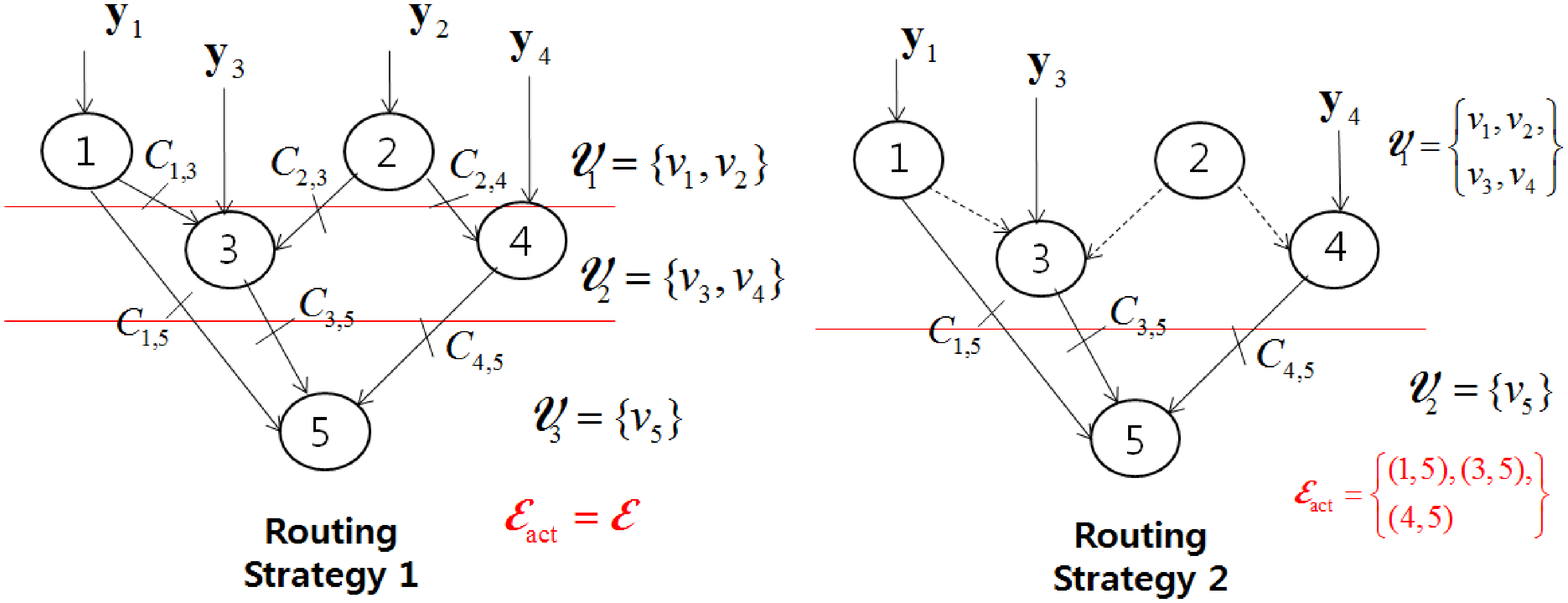}

\caption{{\footnotesize{\label{fig:example-routing}Two different routing schemes
for the same backhaul network with $N_{R}=4$ RUs (dashed arrows represent
inactive edges).}}}
\end{figure}

As discussed in Sec. \ref{sec:Introduction}, we will consider different
strategies for the transmission of the RUs' baseband received signals
to the CU on the backhaul network. For all schemes, routing from the
RUs to the CU can be described as detailed in this subsection following
similar treatments in \cite{Goela-Gastpar}\cite{Healy-Nikolov}. To
this end, we fix an ordered partition of the set $\mathcal{V}$, which
includes the RUs and the CU, into layers $\mathcal{V}_{1},\ldots,\mathcal{V}_{L}$,
so that $\mathcal{V}=\bigcup_{l=1}^{L}\mathcal{V}_{l}$ and $\mathcal{V}_{m}\bigcap\mathcal{V}_{l}=\textrm{�}$
for $m\neq l$ with $N_{R}+1\in\mathcal{V}_{L}$. Each partition gives
rise to a specific routing schedule, as discussed next.

Given a partition $\mathcal{V}_{1},\ldots,\mathcal{V}_{L}$, we consider
as active, and hence available for routing, only the edges, i.e.,
the backhaul links, that connect nodes belonging to successive layers.
More precisely, we define the set $\mathcal{E}_{\mathrm{act}}$ of
the active edges as
\begin{equation}
\mathcal{E}_{\mathrm{act}}=\left\{ e\in\mathcal{E}|\mathrm{tail}(e)\in\mathcal{V}_{l}\,\mathrm{and}\,\mathrm{head}(e)\in\mathcal{V}_{k}\,\,\mathrm{with}\,\, l<k\right\} .\label{eq:active-edge}
\end{equation}
Moreover, we define as $\Gamma_{I}(i)=\{e_{1}^{i},\ldots,e_{|\Gamma_{I}(i)|}^{i}\}$
and $\Gamma_{O}(i)$ the sets of active edges that end or originate
at node $i$, respectively. In other words, we have $\Gamma_{I}(i)=\{e\in\mathcal{E}_{\mathrm{act}}|\mathrm{head}(e)=i\}$
and $\Gamma_{O}(i)=\{e\in\mathcal{E}_{\mathrm{act}}|\mathrm{tail}(e)=i\}$.
The set of nodes that do not have any incoming active edge is denoted
by $\mathcal{S}=\{i\in\mathcal{V}|\Gamma_{I}(i)=\textrm{�}\}$.

A given ordered partition $\mathcal{V}_{1},\ldots,\mathcal{V}_{L}$
defines a routing strategy as follows. Each node $i$ in the first
layer, i.e., with $i\in\mathcal{V}_{1}$, transmits on the active
backhaul links $e\in\Gamma_{O}(i)$ to the nodes in the next layers
$\mathcal{V}_{l}$, $l>1$. The nodes in the second layer $\mathcal{V}_{2}$
wait until all the nodes in the same layer receive from the connected
nodes in $\mathcal{V}_{1}$ and then transmit on the active backhaul
links to the nodes in the next layers $\mathcal{V}_{l}$ with $l>2$.
In general, the nodes in each layer $\mathcal{V}_{l}$ wait for all
the nodes in the same layer to receive from the previous layers $\mathcal{V}_{1},\ldots,\mathcal{V}_{l-1}$
and then transmit on the active backhaul links to the nodes in the
next layers $\mathcal{V}_{l+1},\ldots,\mathcal{V}_{L}$.

Fig. \ref{fig:example-routing} presents two different routing examples
for a backhaul network with $N_{R}=4$ RUs. For routing strategy 1
in the figure, the partition is defined as $\mathcal{V}_{1}=\{1,2\},\mathcal{V}_{2}=\{3,4\},\mathcal{V}_{3}=\{5\}$,
and, as a result, all edges in $\mathcal{E}$ are active, i.e., $\mathcal{E}_{\mathrm{act}}=\mathcal{E}$.
Instead, with routing strategy 2, we have the partition $\mathcal{V}_{1}=\{1,2,3,4\},\mathcal{V}_{2}=\{5\}$
and thus only edges $(1,5)$, $(3,5)$ and $(4,5)$ are active, i.e.,
$\mathcal{E}_{\mathrm{act}}=\{(1,5),(3,5),(4,5)\}$. Note that, with
this strategy, node $2$ does not contribute to the operation of the
network.

\begin{remark}\label{rem:existence-MP}Using classical results in
graph theory, it follows that, for a given directed acyclic graph
$\mathcal{G}=(\mathcal{V},\mathcal{E})$, there always exists a partition
$\mathcal{V}_{1},\ldots,\mathcal{V}_{L}$ that leads to activate all
edges, i.e., to have $\mathcal{E}_{\mathrm{act}}=\mathcal{E}$ (see,
e.g., \cite[Sec. 2.1]{Wainwright}).\end{remark}

We now discuss how the choice of the routing strategy and the the
tolerated delay for communication from the RUs to the CU affect the
use of the capacity of each backhaul link. To start, for a given set
$\mathcal{E}_{\mathrm{act}}$ of active edges, we define as $D_{i}$
the number of edges in the longest path connecting the node $i$ to
the CU $N_{R}+1$. For example, in Fig. \ref{fig:example-routing},
we have $(D_{1},D_{2},D_{3},D_{4})=(2,2,1,1)$ and $(D_{1},D_{3},D_{4})=(1,1,1)$
for routing strategies 1 and 2, respectively. Then, we define the
\textit{depth} $D$ of a routing strategy defined by the partition
$\mathcal{V}_{1},\ldots,\mathcal{V}_{L}$ as $D=\max_{i\in\mathcal{S}}D_{i}$.

Define as $T$ the maximum delay allowed for transmission of the received
baseband signals from the RUs to the CU. We normalize $T$ by the
duration of the transmission on the uplink, so that $T=1$ means that
the delay allowed for transmission on the backhaul network equals
the duration of the uplink transmission. Assuming for simplicity that
each active backhaul link is used for the same amount of time, we
then obtain that each active edge is used only for a period equal
to $T/D$ uplink slots. Therefore, the \textit{effective backhaul
capacity} $\tilde{C}_{e}$ used on an edge $e\in\mathcal{E}_{\mathrm{act}}$,
i.e., the number of bits per channel use of the uplink that are transmitted
on a given active edge $e$, equals $\tilde{C}_{e}=C_{e}\cdot T/D$.
For instance, if $T=D$, and hence a delay equal to the depth of the
routing strategy is tolerated, then we have $\tilde{C}_{e}=C_{e}$.
In this case, in fact, each backhaul link can be activated for the
time duration equal to the wireless uplink transmission block.

\section{Multiplex-and-Forward\label{sec:Forward-without-Decompress}}

\begin{figure}
\centering\includegraphics[width=14cm,height=6cm]{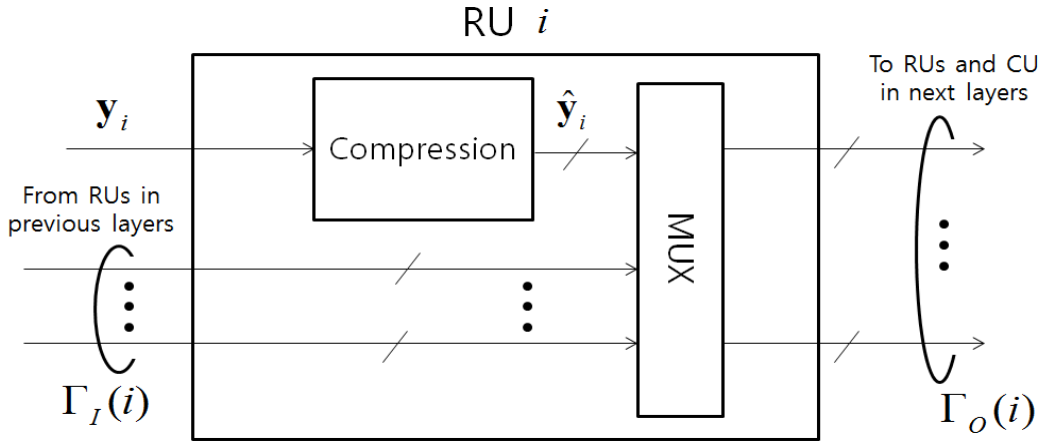}

\caption{{\footnotesize{\label{fig:system-FW}Illustration of the operation
at a RU $i$ in the ``Multiplex-and-Forward'' (MF) scheme studied
in Sec. \ref{sec:Forward-without-Decompress} (Plain arrows ``$\rightarrow$''
indicate baseband signals, while broken arrows ``$\nrightarrow$''
denote bit streams).}}}
\end{figure}

In this section, we present a reference scheme, which we refer to
as Multiplex-and-Forward (MF). In this scheme, as illustrated in Fig.
\ref{fig:system-FW}, each RU $i$ performs compression of its received
baseband signal $\mathbf{y}_{i}$ using a given quantization codebook
and then simply multiplexes the bit streams received from the previous
layers and its compressed signal without any further processing. Specifically,
each RU $i$ transmits on each of its outgoing backhaul links in $\Gamma_{O}(i)$
the bits describing the compressed baseband signal $\hat{\mathbf{y}}_{i}$
within the used quantization codebook along with the bit streams received
from the previous layers. With this scheme, we hence only need to
optimize the compression strategy used to produce $\hat{\mathbf{y}}_{i}$
and the allocation of the backhaul capacity among the received bit
streams and the compressed signal $\hat{\mathbf{y}}_{i}$.

In order to formulate this problem, we define as $f_{e}^{i}\geq0$
the rate (in bits per channel use of the uplink) used to convey the
compressed signal of RU $i$ on edge $e$ for $(i,e)\in\mathcal{N}_{R}\times\mathcal{E}_{\mathrm{act}}$.
By the definition of the routing scheme, we have the following constraints
on the flow variables $f_{e}^{i}$
\begin{align}
f_{e}^{i}\geq & R_{i},\,\,\mathrm{for}\,\, i\in\mathcal{N}_{R}\,\,\mathrm{and}\,\, e\in\Gamma_{O}(i),\label{eq:compression-rate-outgoing}\\
\sum_{e\in\Gamma_{I}(N_{R}+1)}f_{e}^{i}\geq & R_{i},\,\,\mathrm{for}\,\, i\in\mathcal{N}_{R},\label{eq:compression-rate-incoming}\\
\sum_{i\in\mathcal{N}_{R}}f_{e}^{i}\leq & \tilde{C}_{e},\,\,\mathrm{for}\,\, e\in\mathcal{E}_{\mathrm{act}},\label{eq:edge-conditions}\\
\mathrm{and}\,\,\sum_{e\in\Gamma_{I}(j)}f_{e}^{i}\geq & \sum_{e\in\Gamma_{O}(j)}f_{e}^{i},\,\,\mathrm{for}\,\,(j,i)\in\mathcal{N}_{R}\times\mathcal{N}_{R},\label{eq:flow-conversion-rule}
\end{align}
where $R_{i}$ represents the rate at which RU $i$ compresses its
baseband signal $\mathbf{y}_{i}$. This information must be sent on
all the outgoing links $\Gamma_{O}(i)$ as per (\ref{eq:compression-rate-outgoing}).
The condition (\ref{eq:compression-rate-incoming}) guarantees that
the CU $N_{R}+1$ receives sufficient information to be able to decompress,
and the constraints (\ref{eq:edge-conditions}) impose that the sum
of the capacities $\{f_{e}^{i}\}_{i\in\mathcal{N}_{R}}$ passing through
an edge $e$ does not exceed the effective capacity $\tilde{C}_{e}$.
The last condition (\ref{eq:flow-conversion-rule}) represents the
flow conservation rule at each RU $j\in\mathcal{N}_{R}$.

In order to describe the relationship between the rate $R_{i}$ and
the fidelity of the compressed signal $\hat{\mathbf{y}}_{i}$, we
use standard rate distortion theoretic arguments (e.g., \cite[Ch. 3]{ElGamal-Kim}).
Specifically, as in, e.g., \cite{dCoso}\cite{Chechik-et-al}\cite{Tian-Chen},
we assume a Gaussian quantization noise (without claim of optimality),
so that the signal $\mathbf{\hat{\boldsymbol{\mathrm{y}}}}_{i}$ is
given by%
\footnote{As discussed in \cite{dCoso}, the model (\ref{eq:test-channel-FW})
is as general as the model $\mathbf{\hat{\boldsymbol{\mathrm{y}}}}_{i}=\mathbf{L}_{i}\mathbf{y}_{i}+\mathbf{q}_{i}$
that contains a linear processing $\mathbf{L}_{i}$ prior to compression.%
}
\begin{equation}
\mathbf{\hat{\boldsymbol{\mathrm{y}}}}_{i}=\mathbf{y}_{i}+\mathbf{q}_{i}.\label{eq:test-channel-FW}
\end{equation}
In (\ref{eq:test-channel-FW}), $\mathbf{q}_{i}$ represents the quantization
noise, which is distributed as $\mathcal{CN}(\mathbf{0},\mathbf{\Omega}_{i})$
and is independent of $\mathbf{y}_{i}$. From rate-distortion theory,
the compressed signal $\mathbf{\hat{\boldsymbol{\mathrm{y}}}}_{i}$
in (\ref{eq:test-channel-FW}) can be obtained at the output of the
compressor if the rate $R_{i}$ satisfies the inequality \cite[Ch. 3]{ElGamal-Kim}
\begin{align}
g_{i}^{\mathrm{MF}}\left(\{\mathbf{\Omega}_{i}\}_{i\in\mathcal{M}}\right)\triangleq & I\left(\mathbf{y}_{i};\mathbf{\hat{\boldsymbol{\mathrm{y}}}}_{i}\right)\label{eq:backhaul-constraint-FW}\\
= & \log\det\left(\mathbf{\Omega}_{i}+\mathbf{\Sigma}_{\mathbf{y}_{i}}\right)-\log\det\left(\mathbf{\Omega}_{i}\right)\leq R_{i}.\nonumber
\end{align}

Based on the discussion above, as long as the constraints (\ref{eq:compression-rate-outgoing})-(\ref{eq:flow-conversion-rule})
and (\ref{eq:backhaul-constraint-FW}) are satisfied, the CU is able
to recover the signals $\hat{\mathbf{y}}_{i}$, $i\in\mathcal{N}_{R}$,
and an achievable sum-rate $R_{\mathrm{sum}}$ between the MSs and
the CU is given as
\begin{align}
R_{\mathrm{sum}}= & I\left(\mathbf{x};\{\mathbf{\hat{\boldsymbol{\mathrm{y}}}}_{i}\}_{i\in\mathcal{N}_{R}}\right)=f^{\mathrm{MF}}\left(\{\mathbf{\Omega}_{i}\}_{i\in\mathcal{N}_{R}}\right)\label{eq:sum-rate-FW}\\
\triangleq & \log\det\left(\mathbf{H}\mathbf{\Sigma}_{\mathbf{x}}\mathbf{H}^{\dagger}+\mathbf{I}+\mathbf{\Omega}\right)-\log\det\left(\mathbf{I}+\mathbf{\Omega}\right),\nonumber
\end{align}
with the definition $\mathbf{\Omega}=\mathrm{diag}(\mathbf{\Omega}_{1},\ldots,\mathbf{\Omega}_{N_{R}})$.

\subsection{Problem Definition and Optimization}

For a given routing strategy defined by the partition $\mathcal{V}_{1},\ldots,\mathcal{V}_{L}$,
we aim at optimizing the compression strategies $\{\mathbf{\Omega}_{i}\}_{i\in\mathcal{N}_{R}}$
and the flow variables $\{f_{e}^{i}\}_{i\in\mathcal{N}_{R},e\in\mathcal{E}_{\mathrm{act}}}$
with the goal of maximizing the sum-rate $R_{\mathrm{sum}}$ in (\ref{eq:sum-rate-FW})
subject to the constraints (\ref{eq:compression-rate-outgoing})-(\ref{eq:flow-conversion-rule})
and (\ref{eq:backhaul-constraint-FW}). This problem is stated as\begin{subequations}\label{eq:problem-FW}
\begin{align}
\underset{\begin{array}{c}
\{\mathbf{\Omega}_{i}\succeq\mathbf{0},R_{i}\geq0\}_{i\in\mathcal{N}_{R}},\\
\{f_{e}^{i}\geq0\}_{i\in\mathcal{N}_{R},e\in\mathcal{E}_{\mathrm{act}}}
\end{array}}{\mathrm{maximize}} & f^{\mathrm{MF}}\left(\{\mathbf{\Omega}_{i}\}_{i\in\mathcal{N}_{R}}\right)\label{eq:problem-FW-objective}\\
\mathrm{s.t.}\,\,\,\,\,\, & g_{i}^{\mathrm{MF}}\left(\{\mathbf{\Omega}_{i}\}_{i\in\mathcal{N}_{R}}\right)\leq R_{i},\,\,\mathrm{for}\,\, i\in\mathcal{N}_{R},\label{eq:problem-FW-compression}\\
 & (\ref{eq:compression-rate-outgoing})-(\ref{eq:flow-conversion-rule}).\label{eq:problem-FW-remainder}
\end{align}
\end{subequations}We note that the optimization (\ref{eq:problem-FW})
requires full CSI.

The problem (\ref{eq:problem-FW}) with respect to the variables $\{\mathbf{\Omega}_{i}\}_{i\in\mathcal{N}_{R}}$
and $\{f_{e}^{i}\geq0\}_{i\in\mathcal{N}_{R},e\in\mathcal{E}_{\mathrm{act}}}$
is a difference-of-convex problem, which is a subclass of non-convex
problems with desirable properties \cite{Beck}. The problem is a
difference-of-convex problem because the functions $f^{\mathrm{MF}}(\{\mathbf{\Omega}_{i}\}_{i\in\mathcal{N}_{R}})$
and $g_{i}^{\mathrm{MF}}(\{\mathbf{\Omega}_{i}\}_{i\in\mathcal{N}_{R}})$
can be written as the difference of convex functions and all other
constraints in (\ref{eq:problem-FW-remainder}) are linear. For difference-of-convex
problems, the Majorization and Minimization (MM) algorithm provides
an iterative procedure that is known to converge to a stationary point
of the problem (see, e.g., \cite{Beck}). The detailed algorithm is
described in Algorithm 1, where we have defined for brevity the functions
$\tilde{f}^{\mathrm{MF}}(\{\mathbf{\Omega}_{i}^{(t+1)},\mathbf{\Omega}_{i}^{(t)}\}_{i\in\mathcal{N}_{R}})$
and $\tilde{g}_{i}^{\mathrm{MF}}(\{\mathbf{\Omega}_{i}^{(t+1)},\mathbf{\Omega}_{i}^{(t)}\}_{i\in\mathcal{N}_{R}})$
as
\begin{align}
\tilde{f}^{\mathrm{MF}}\left(\{\mathbf{\Omega}_{i}^{(t+1)},\mathbf{\Omega}_{i}^{(t)}\}_{i\in\mathcal{N}_{R}}\right)\triangleq & \log\det\left(\mathbf{H}\mathbf{\Sigma}_{\mathbf{x}}\mathbf{H}^{\dagger}+\mathbf{I}+\mathbf{\Omega}^{(t+1)}\right)\\
- & \varphi(\mathbf{I}+\mathbf{\Omega}^{(t+1)},\mathbf{I}+\mathbf{\Omega}^{(t)})\nonumber \\
\mathrm{and}\,\,\tilde{g}_{i}^{\mathrm{MF}}\left(\{\mathbf{\Omega}_{i}^{(t+1)},\mathbf{\Omega}_{i}^{(t)}\}_{i\in\mathcal{N}_{R}}\right)\triangleq & \varphi(\mathbf{\Omega}_{i}^{(t+1)}+\mathbf{\Sigma}_{\mathbf{y}_{i}},\mathbf{\Omega}_{i}^{(t)}+\mathbf{\Sigma}_{\mathbf{y}_{i}})-\log\det\left(\mathbf{\Omega}_{i}^{(t+1)}\right),
\end{align}
with the function $\varphi(\mathbf{X},\mathbf{Y})$ given as
\begin{equation}
\varphi(\mathbf{X},\mathbf{Y})\triangleq\log\det\left(\mathbf{Y}\right)+\frac{1}{\ln2}\mathrm{tr}\left(\mathbf{Y}^{-1}\left(\mathbf{X}-\mathbf{Y}\right)\right).
\end{equation}

\begin{algorithm}
\caption{MF: MM Algorithm for problem (\ref{eq:problem-FW})}

1. Initialize the matrices $\{\mathbf{\Omega}_{i}^{(1)}\}_{i\in\mathcal{N}_{R}}$
to arbitrary feasible positive semidefinite matrices for problem (\ref{eq:problem-FW})
and set $t=1$.

2. Update the matrices $\{\mathbf{\Omega}_{i}^{(t+1)}\}_{i\in\mathcal{N}_{R}}$
and variables $\{f_{e}^{i}\geq0\}_{i\in\mathcal{N}_{R},e\in\mathcal{E}_{\mathrm{act}}}$
as a solution of the following convex problem:
\begin{align}
\underset{\begin{array}{c}
\{\mathbf{\Omega}_{i}^{(t+1)}\succeq\mathbf{0},R_{i}\geq0\}_{i\in\mathcal{N}_{R}},\\
\{f_{e}^{i}\geq0\}_{i\in\mathcal{N}_{R},e\in\mathcal{E}_{\mathrm{act}}}
\end{array}}{\mathrm{maximize}} & \tilde{f}^{\mathrm{MF}}\left(\{\mathbf{\Omega}_{i}^{(t+1)},\mathbf{\Omega}_{i}^{(t)}\}_{i\in\mathcal{N}_{R}}\right)\label{eq:problem-convexified-centralized-1}\\
\mathrm{s.t.}\,\,\,\,\,\,\,\,\, & \tilde{g}_{i}^{\mathrm{MF}}\left(\{\mathbf{\Omega}_{i}^{(t+1)},\mathbf{\Omega}_{i}^{(t)}\}_{i\in\mathcal{N}_{R}}\right)\leq R_{i},\,\,\mathrm{for}\,\, i\in\mathcal{N}_{R},\nonumber \\
 & (\ref{eq:compression-rate-outgoing})-(\ref{eq:flow-conversion-rule}).\nonumber
\end{align}

3. Stop if a convergence criterion is satisfied. Otherwise, set $t\leftarrow t+1$
and go back to Step 2.
\end{algorithm}

\section{Decompress-Process-and-Recompress\label{sec:Decompress-and-Recompress}}

\begin{figure}
\centering\includegraphics[width=14cm,height=6cm]{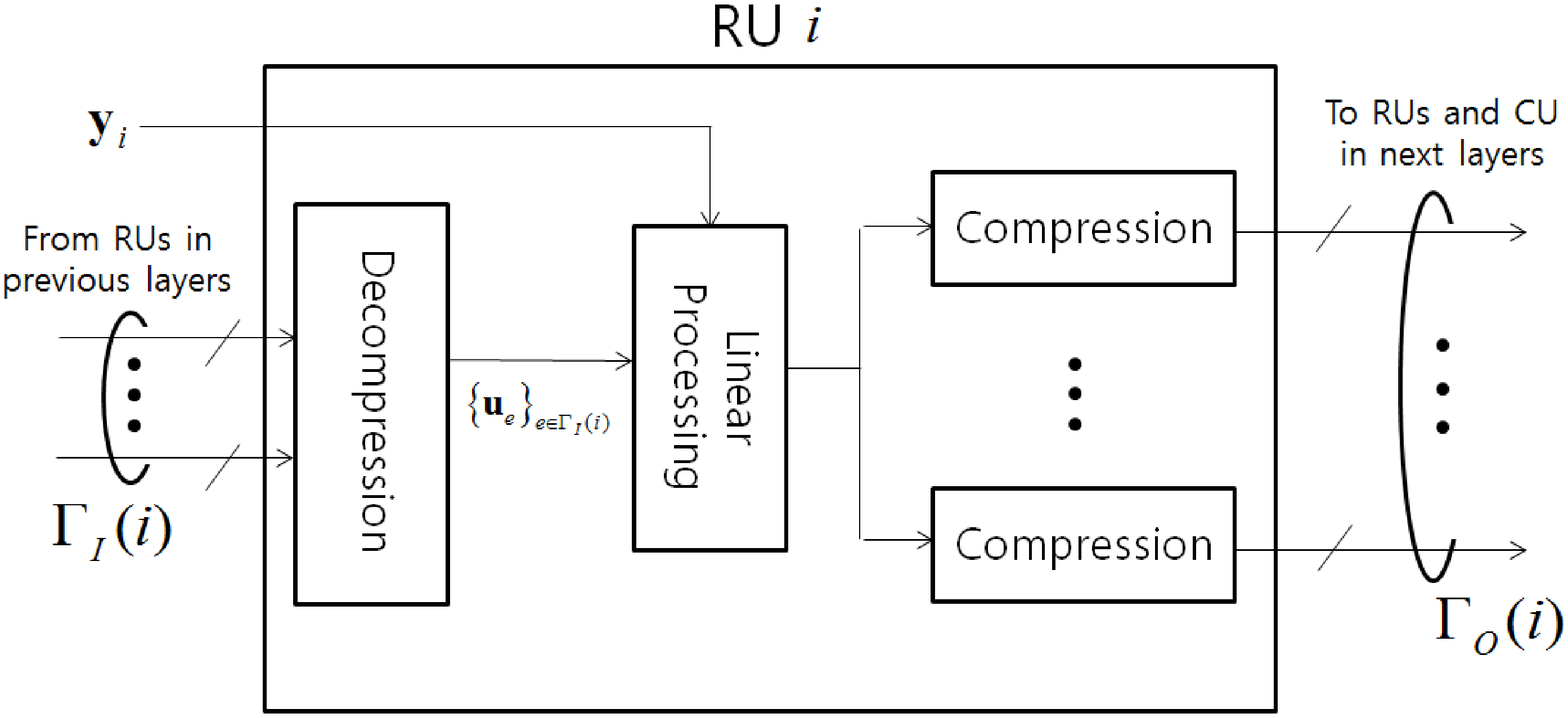}

\caption{{\footnotesize{\label{fig:system-DR}Illustration of the operation
at a RU $i$ in the ``Decompress-Process-and-Recompress'' (DPR)
scheme studied in Sec. \ref{sec:Decompress-and-Recompress} (Plain
arrows ``$\rightarrow$'' indicate baseband signals, while broken
arrows ``$\nrightarrow$'' denote bit streams).}}}
\end{figure}

The MF backhaul strategy studied in the previous section may incur
a significant performance degradation when the RUs have a sufficiently
large number of incoming edges. In fact, in this case, the bit rate
obtained by multiplexing the signals received from the RUs in the
previous layers is large and the backhaul capacity constraints may
impose a critical performance bottleneck. In this section, we introduce
a scheme that attempts to solve this problem via decompression at
each RU and linear in-network processing of the decompressed signals
and of the locally received signal. The key idea is that the processing
step can reduce redundancy by properly combining the available (compressed)
received signals. On the flip side, the processed signals need to
be recompressed before they can be sent on the backhaul links. As
discussed in \cite{Cuff-et-al} in the context of a cascade source
coding problem, this recompression step introduces further distortion.
The effect of this distortion must thus be counterbalanced by the
advantages of in-network processing in order to make the strategy
preferable to MF.

We now detail the DPR scheme and analyze its performance. As shown
in Fig. \ref{fig:system-DR}, each RU $i$ first decompresses the
signals $\mathbf{u}_{e^{\prime}}$ received on its incoming edges
$e^{\prime}\in\Gamma_{I}(i)$. Then, for each outgoing edge $e\in\Gamma_{O}(i)$,
it processes the vector $\mathbf{r}_{i}$ that includes the decompressed
signals $\mathbf{u}_{e^{\prime}}$ for all edges $e^{\prime}\in\Gamma_{I}(i)$
and the received baseband signal $\mathbf{y}_{i}$, namely
\begin{equation}
\mathbf{r}_{i}=[\mathbf{y}_{i};\mathbf{u}_{e_{1}^{i}};\cdots;\mathbf{u}_{e_{\left|\Gamma_{I}(i)\right|}^{i}}],\label{eq:received-signal-aggregated}
\end{equation}
via a linear processing matrix $\mathbf{L}_{e}$. This produces a
processed signal $\mathbf{L}_{e}\mathbf{r}_{i}$ for all outgoing
edges $e\in\Gamma_{O}(i)$. We assume here that matrix $\mathbf{L}_{e}$
is square and study the issue of dimensionality reduction via the
use of ``wide'' matrices $\mathbf{L}_{e}$ in Sec. \ref{sub:Limited-Rank-Processing}.
Note that the matrix $\mathbf{L}_{e}$ can be written as
\begin{equation}
\mathbf{L}_{e}=[\mathbf{L}_{e}^{\mathrm{rx}}\,\mathbf{L}_{e}^{e_{1}^{i}}\,\cdots\,\mathbf{L}_{e}^{e_{\left|\Gamma_{I}(i)\right|}^{i}}],
\end{equation}
where, by (\ref{eq:received-signal-aggregated}), the matrices $\mathbf{L}_{e}^{\mathrm{rx}}\in\mathbb{C}^{d_{e}\times n_{R,i}}$
and $\mathbf{L}_{e}^{e_{j}^{i}}\in\mathbb{C}^{d_{e}\times d_{e_{j}^{i}}}$
multiply the signals $\mathbf{y}_{i}$ and $\mathbf{u}_{e_{j}^{i}}$,
respectively, for $j\in\{1,\ldots,\left|\Gamma_{I}(i)\right|\}$.
Finally, RU $i$ compresses the processed signal $\mathbf{L}_{e}\mathbf{r}_{i}$
at a rate of $\tilde{C}_{e}$ bits per channel use to produce the
output signal $\mathbf{u}_{e}$ to be sent on the outgoing active
edge $e\in\Gamma_{O}(i)$.

As in the previous section, we leverage standard rate-distortion theory
arguments to model compression and we adopt (without claim of optimality)
a Gaussian quantization noise, so that the signal $\mathbf{u}_{e}$
is given by
\begin{equation}
\mathbf{u}_{e}=\mathbf{L}_{e}\mathbf{r}_{i}+\mathbf{q}_{e},\label{eq:Gaussian-test-channel}
\end{equation}
with quantization noise $\mathbf{q}_{e}$ being distributed as $\mathcal{CN}(\mathbf{0},\mathbf{\Omega}_{e})$.
Moreover, we assume that the signals $\mathbf{u}_{e}$ and $\mathbf{u}_{e^{\prime}}$,
to be delivered on different outgoing edges $e$ and $e^{\prime}$
with $e\neq e^{\prime}\in\Gamma_{O}(i)$, are quantized with independent
codebooks so that the quantization noises $\mathbf{q}_{e}$ and $\mathbf{q}_{e^{\prime}}$
are independent of each other. Similar to (\ref{eq:backhaul-constraint-FW}),
the signal $\mathbf{u}_{e}$ can be reliably transmitted to RU $\mathrm{head}(e)$
if the condition
\begin{align}
g_{e}^{\mathrm{DPR}}\left(\{\mathbf{L}_{e},\mathbf{\Omega}_{e}\}_{e\in\mathcal{E}_{\mathrm{act}}}\right)\triangleq & I\left(\mathbf{r}_{i};\mathbf{u}_{e}\right)\label{eq:backhaul-constraint-DR}\\
= & \log\det\left(\mathbf{\Omega}_{e}+\mathbf{L}_{e}\mathbf{\Sigma}_{\mathbf{r}_{i}}\mathbf{L}_{e}^{\dagger}\right)-\log\det\left(\mathbf{\Omega}_{e}\right)\leq\tilde{C}_{e}\nonumber
\end{align}
is satisfied.

The CU performs joint decoding of the messages of all MSs based on
the received signal $\mathbf{r}_{N_{R}+1}$, which can be written,
similar to (\ref{eq:received-signal-aggregated}), as
\begin{equation}
\mathbf{r}_{N_{R}+1}=[\mathbf{u}_{e_{1}^{N_{R}+1}};\cdots;\mathbf{u}_{e_{\left|\Gamma_{I}(N_{R}+1)\right|}^{N_{R}+1}}].\label{eq:received-signal-CU}
\end{equation}
As a result, the sum-rate
\begin{equation}
R_{\mathrm{sum}}=I\left(\mathbf{x};\mathbf{r}_{N_{R}+1}\right)\label{eq:sum-rate-DR}
\end{equation}
is achievable between the MSs and the CU. The sum-rate (\ref{eq:sum-rate-DR})
is characterized in the following lemma.

\begin{lemma}\label{lem:input-output}For any given routing strategy
defined by the partition $\mathcal{V}_{1},\ldots,\mathcal{V}_{L}$
with the active edges $\mathcal{E}_{\mathrm{act}}=\{e_{1},\ldots,e_{|\mathcal{E}_{\mathrm{act}}|}\}$,
the sum-rate $R_{\mathrm{sum}}$ in (\ref{eq:sum-rate-DR}) is given
by
\begin{align}
R_{\mathrm{sum}}= & f^{\mathrm{DPR}}\left(\{\mathbf{L}_{e},\mathbf{\Omega}_{e}\}_{e\in\mathcal{E}_{\mathrm{act}}}\right)\label{eq:sum-rate-computed}\\
\triangleq & \log\det\left(\mathbf{T}\mathbf{H}\mathbf{\Sigma}_{\mathbf{x}}\mathbf{H}^{\dagger}\mathbf{T}^{\dagger}+\mathbf{T}\mathbf{T}^{\dagger}+\tilde{\mathbf{T}}\mathbf{\Omega}\tilde{\mathbf{T}}^{\dagger}\right)-\log\det\left(\mathbf{T}\mathbf{T}^{\dagger}+\tilde{\mathbf{T}}\mathbf{\Omega}\tilde{\mathbf{T}}^{\dagger}\right),\nonumber
\end{align}
where $\mathbf{\Omega}=\mathrm{diag}(\mathbf{\Omega}_{e_{1}},\ldots,\mathbf{\Omega}_{e_{|\mathcal{E}_{\mathrm{act}}|}})$
and the matrices $\mathbf{T}$ and $\tilde{\mathbf{T}}$ are defined
as
\begin{equation}
\mathbf{T}=\mathbf{C}\left(\mathbf{I}-\mathbf{F}\right)^{-1}\mathbf{E}\,\,\mathrm{and}\,\,\tilde{\mathbf{T}}=\mathbf{C}\left(\mathbf{I}-\mathbf{F}\right)^{-1},\label{eq:transfer-matrices-DR}
\end{equation}
with
\begin{align}
\mathbf{C} & =\left[\begin{array}{ccc}
\mathbf{C}_{e_{1}^{N_{R}+1},e_{1}} & \cdots & \mathbf{C}_{e_{1}^{N_{R}+1},e_{\left|\mathcal{E}_{\mathrm{act}}\right|}}\\
\vdots & \ddots & \vdots\\
\mathbf{C}_{e_{\left|\Gamma_{I}(N_{R}+1)\right|}^{N_{R}+1},e_{1}} & \cdots & \mathbf{C}_{e_{\left|\Gamma_{I}(N_{R}+1)\right|}^{N_{R}+1},e_{\left|\mathcal{E}_{\mathrm{act}}\right|}}
\end{array}\right],\\
\mathbf{F} & =\left[\begin{array}{ccc}
\mathbf{F}_{e_{1},e_{1}} & \cdots & \mathbf{F}_{e_{1},e_{\left|\mathcal{E}_{\mathrm{act}}\right|}}\\
\vdots & \ddots & \vdots\\
\mathbf{F}_{e_{\left|\mathcal{E}_{\mathrm{act}}\right|},e_{1}} & \cdots & \mathbf{F}_{e_{\left|\mathcal{E}_{\mathrm{act}}\right|},e_{\left|\mathcal{E}_{\mathrm{act}}\right|}}
\end{array}\right],\\
\mathrm{and}\,\,\mathbf{E} & =\left[\begin{array}{ccc}
\mathbf{E}_{e_{1},1} & \cdots & \mathbf{E}_{e_{1},N_{R}}\\
\vdots & \ddots & \vdots\\
\mathbf{E}_{e_{\left|\mathcal{E}_{\mathrm{act}}\right|},1} & \cdots & \mathbf{E}_{e_{\left|\mathcal{E}_{\mathrm{act}}\right|},N_{R}}
\end{array}\right],
\end{align}
where
\begin{align}
\mathbf{C}_{e,e^{\prime}} & =\begin{cases}
\mathbf{I}, & \mathrm{if}\, e=e^{\prime}\\
\mathbf{0}, & \mathrm{otherwise}
\end{cases},\\
\mathbf{F}_{e,e^{\prime}} & =\begin{cases}
\mathbf{L}_{e}^{e^{\prime}}, & \mathrm{if}\,\mathrm{tail}(e)=\mathrm{head}(e^{\prime})\\
\mathbf{0}, & \mathrm{otherwise}
\end{cases},\\
\mathrm{and}\,\,\mathbf{E}_{e,j} & =\begin{cases}
\mathbf{L}_{e}^{\mathrm{rx}}, & \mathrm{if}\,\mathrm{tail}(e)=j\\
\mathbf{0}, & \mathrm{otherwise}
\end{cases}.
\end{align}
\end{lemma}\textit{Proof:} The result follows by noting that the
signal $\mathbf{r}_{N_{R}+1}$ in (\ref{eq:received-signal-CU}) received
by the CU $N_{R}+1$ can be written as
\begin{equation}
\mathbf{r}_{N_{R}+1}=\mathbf{T}\mathbf{y}+\tilde{\mathbf{T}}\mathbf{q},\label{eq:input-output}
\end{equation}
with the quantization noise vector $\mathbf{q}=[\mathbf{q}_{e_{1}};\cdots;\mathbf{q}_{e_{\left|\mathcal{E}_{\mathrm{act}}\right|}}]\sim\mathcal{CN}(\mathbf{0},\mathbf{\Omega})$.
This can be proved by identifying the state-space equations and linear
transfer functions as done in \cite[Sec. III-A]{Goela-Gastpar}. $\square$

\subsection{Problem Definition and Optimization\label{sub:Problem-Definition-DR}}

For a given routing strategy defined by the partition $\mathcal{V}_{1},\ldots,\mathcal{V}_{L}$,
we are interested in tackling the problem of maximizing the sum-rate
(\ref{eq:sum-rate-computed}) over the variables $\{\mathbf{L}_{e},\mathbf{\Omega}_{e}\}_{e\in\mathcal{E}_{\mathrm{act}}}$.
This problem can be stated as\begin{subequations}\label{eq:problem-DR}
\begin{align}
\underset{\{\mathbf{L}_{e},\mathbf{\Omega}_{e}\succeq\mathbf{0}\}_{e\in\mathcal{E}_{\mathrm{act}}}}{\mathrm{maximize}} & f^{\mathrm{DPR}}\left(\{\mathbf{L}_{e},\mathbf{\Omega}_{e}\}_{e\in\mathcal{E}_{\mathrm{act}}}\right)\label{eq:problem-objective}\\
\mathrm{s.t.}\,\,\, & g_{e}^{\mathrm{DPR}}\left(\{\mathbf{L}_{e},\mathbf{\Omega}_{e}\}_{e\in\mathcal{E}_{\mathrm{act}}}\right)\leq\tilde{C}_{e},\,\,\mathrm{for}\, e\in\mathcal{E}_{\mathrm{act}}.\label{eq:problem-backhaul}
\end{align}
\end{subequations}

We now discuss the optimization (\ref{eq:problem-DR}) of the compression
strategies $\{\mathbf{L}_{e},\mathbf{\Omega}_{e}\}_{e\in\mathcal{E}_{\mathrm{act}}}$
under the assumption that full CSI is available at the optimizing
unit. Decentralized optimization based on local CSI at each node will
be studied in Sec. \ref{sec:Decentralized-Optimization}. The following
proposition shows that, under the stated assumptions, we can fix the
linear transformations $\mathbf{L}_{e}$ to be equal to an identity
matrix, i.e., $\mathbf{L}_{e}=\mathbf{I}$ for all $e\in\mathcal{E}_{\mathrm{act}}$,
without loss of optimality.

\begin{proposition}\label{prop:identity}For any solution $\{\mathbf{L}_{e}^{\prime},\mathbf{\Omega}_{e}^{\prime}\}_{e\in\mathcal{E}_{\mathrm{act}}}$
of problem (\ref{eq:problem-DR}), there exists another equivalent
solution $\{\mathbf{L}_{e}^{\prime\prime},\mathbf{\Omega}_{e}^{\prime\prime}\}_{e\in\mathcal{E}_{\mathrm{act}}}$
with $\mathbf{L}_{e}^{\prime\prime}=\mathbf{I}$, in the sense that
$f^{\mathrm{DPR}}(\{\mathbf{L}_{e}^{\prime},\mathbf{\Omega}_{e}^{\prime}\}_{e\in\mathcal{E}_{\mathrm{act}}})=f^{\mathrm{DPR}}(\{\mathbf{L}_{e}^{\prime\prime},\mathbf{\Omega}_{e}^{\prime\prime}\}_{e\in\mathcal{E}_{\mathrm{act}}})$
and $g_{e}^{\mathrm{DPR}}(\{\mathbf{L}_{e}^{\prime},\mathbf{\Omega}_{e}^{\prime}\}_{e\in\mathcal{E}_{\mathrm{act}}})=g_{e}^{\mathrm{DPR}}(\{\mathbf{L}_{e}^{\prime\prime},\mathbf{\Omega}_{e}^{\prime\prime}\}_{e\in\mathcal{E}_{\mathrm{act}}})$
for all $e\in\mathcal{E}_{\mathrm{act}}$. \end{proposition}\textit{Proof:}
See Appendix \ref{appendix:proof-lem-identity}.~~~~~~~~~~~~~~~~~~~~~~~~~~~~~~~~~~~~~~~~~~~~~~~~~~~~~~~~~~~~~~~~~~~~~~~~~~~~~~~~~~~~~~~~~~
$\square$~

Using Proposition \ref{prop:identity}, the problem (\ref{eq:problem-DR})
can be reduced with no loss of optimality to an optimization solely
with respect to the quantization noise covariances $\{\mathbf{\Omega}_{e}\}_{e\in\mathcal{E}_{\mathrm{act}}}$.
The mentioned optimization of (\ref{eq:problem-DR}) with $\mathbf{L}_{e}=\mathbf{I}$,
$e\in\mathcal{E}_{\mathrm{act}}$, can be seen to be a difference-of-convex
problem, as introduced in Sec. \ref{sec:Forward-without-Decompress}.
Therefore, we can apply the MM approach \cite{Beck} to find a stationary
point of the problem as in Sec. \ref{sec:Forward-without-Decompress}.
The derived algorithm, which is referred to as ``DPR-opt'', is described
in Algorithm 2, where we have defined the functions $\tilde{f}^{\mathrm{DPR}}(\{\mathbf{\Omega}_{e}^{(t+1)},\mathbf{\Omega}_{e}^{(t)}\}_{e\in\mathcal{E}_{\mathrm{act}}})$
and $\tilde{g}_{e}^{\mathrm{DPR}}(\{\mathbf{\Omega}_{e}^{(t+1)},\mathbf{\Omega}_{e}^{(t)}\}_{e\in\mathcal{E}_{\mathrm{act}}})$
as
\begin{align}
\tilde{f}^{\mathrm{DPR}}\left(\{\mathbf{\Omega}_{e}^{(t+1)},\mathbf{\Omega}_{e}^{(t)}\}_{e\in\mathcal{E}_{\mathrm{act}}}\right)= & \log\det\left(\mathbf{T}\mathbf{H}\mathbf{\Sigma}_{\mathbf{x}}\mathbf{H}^{\dagger}\mathbf{T}^{\dagger}+\mathbf{T}\mathbf{T}^{\dagger}+\tilde{\mathbf{T}}\mathbf{\Omega}^{(t+1)}\tilde{\mathbf{T}}^{\dagger}\right)\label{eq:function-DR-obj}\\
- & \varphi\left(\mathbf{T}\mathbf{T}^{\dagger}+\tilde{\mathbf{T}}\mathbf{\Omega}^{(t+1)}\tilde{\mathbf{T}}^{\dagger},\mathbf{T}\mathbf{T}^{\dagger}+\tilde{\mathbf{T}}\mathbf{\Omega}^{(t)}\tilde{\mathbf{T}}^{\dagger}\right),\nonumber \\
\mathrm{and}\,\,\tilde{g}_{e}^{\mathrm{DPR}}\left(\{\mathbf{\Omega}_{e}^{(t+1)},\mathbf{\Omega}_{e}^{(t)}\}_{e\in\mathcal{E}_{\mathrm{act}}}\right)= & \varphi\left(\mathbf{\Omega}_{e}^{(t+1)}+\mathbf{L}_{e}\mathbf{\Sigma}_{\mathbf{r}_{i}}^{(t+1)}\mathbf{L}_{e}^{\dagger},\mathbf{\Omega}_{e}^{(t)}+\mathbf{L}_{e}\mathbf{\Sigma}_{\mathbf{r}_{i}}^{(t)}\mathbf{L}_{e}^{\dagger}\right)-\log\det\left(\mathbf{\Omega}_{e}^{(t+1)}\right).\label{eq:function-DR-backhaul}
\end{align}

\begin{algorithm}
\caption{DPR-opt: MM Algorithm for problem (\ref{eq:problem-DR}) with fixed
$\{\mathbf{L}_{e}=\mathbf{I}\}_{e\in\mathcal{E}_{\mathrm{\mathrm{act}}}}$}

1. Initialize the matrices $\{\mathbf{\Omega}_{e}^{(1)}\}_{e\in\mathcal{E}_{\mathrm{act}}}$
to arbitrary feasible positive semidefinite matrices for problem (\ref{eq:problem-DR})
and set $t=1$.

2. Update the matrices $\{\mathbf{\Omega}_{e}^{(t+1)}\}_{e\in\mathcal{E}_{\mathrm{act}}}$
as a solution of the following (convex) problem.
\begin{align}
\underset{\{\mathbf{\Omega}_{e}^{(t+1)}\succeq\mathbf{0}\}_{e\in\mathcal{E}_{\mathrm{act}}}}{\mathrm{maximize}} & \tilde{f}^{\mathrm{DPR}}\left(\{\mathbf{\Omega}_{e}^{(t+1)},\mathbf{\Omega}_{e}^{(t)}\}_{e\in\mathcal{E}_{\mathrm{act}}}\right)\label{eq:problem-convexified-centralized}\\
\mathrm{s.t.}\,\,\,\,\,\,\,\,\, & \tilde{g}_{e}^{\mathrm{DPR}}\left(\{\mathbf{\Omega}_{e}^{(t+1)},\mathbf{\Omega}_{e}^{(t)}\}_{e\in\mathcal{E}_{\mathrm{act}}}\right)\leq\tilde{C}_{e},\,\,\mathrm{for}\, e\in\mathcal{E}_{\mathrm{act}}.\nonumber
\end{align}

3. Stop if a convergence criterion is satisfied. Otherwise, set $t\leftarrow t+1$
and go back to Step 2.
\end{algorithm}

\subsection{Limited-Rank Processing\label{sub:Limited-Rank-Processing}}

When the backhaul network consists of a large number $L$ of layers,
Algorithm 2 may have a prohibitive complexity due to the large dimensionality
of the signal $\mathbf{r}_{i}$ in (\ref{eq:received-signal-aggregated})
to be processed at each RU $i$. The dimension $d_{i}$ of the signal
$\mathbf{r}_{i}$ in (\ref{eq:received-signal-aggregated}) can in
fact be recursively computed as $d_{i}=n_{R,i}+\sum_{e\in\Gamma_{I}(i)}d_{\mathrm{tail}(e)}$.
In order to tackle this problem, we impose a dimensionality constraint
$d_{e}\leq d_{\mathrm{tail}(e)}$ on the active edges $e\in\mathcal{E}_{\mathrm{act}}$.
This is done by constraining the matrices $\mathbf{L}_{e}$ to be
$d_{e}\times d_{\mathrm{tail}(e)}$ rather than the square $d_{\mathrm{tail}(e)}\times d_{\mathrm{tail}(e)}$
matrices $\mathbf{L}_{e}$ considered up to now, where $d_{i}$ can
now be computed recursively as $d_{i}=n_{R,i}+\sum_{e\in\Gamma_{I}(i)}d_{e}$.
Under this constraint, Proposition \ref{prop:identity} does not hold
any more and one must jointly optimize the matrices $\{\mathbf{L}_{e}\}_{e\in\mathcal{E}_{\mathrm{act}}}$
and $\{\mathbf{\Omega}_{e}\}_{e\in\mathcal{E}_{\mathrm{act}}}$. To
this end, we propose a (generally suboptimal) three-step approach,
referred to as ``DPR-rank-$d_{e}$'', as described in Algorithm
3. In this algorithm, at each step, we perform the optimization with
respect to the covariances $\{\mathbf{\Omega}_{e}\}_{e\in\mathcal{E}_{\mathrm{act}}}$
for fixed transforms $\{\mathbf{L}_{e}\}_{e\in\mathcal{E}_{\mathrm{act}}}$
using Algorithm 2, and then update the matrices $\{\mathbf{L}_{e}\}_{e\in\mathcal{E}_{\mathrm{act}}}$
based on the eigenvectors corresponding to the $d_{e}$ smallest eigenvalues
of the obtained matrices $\{\mathbf{\Omega}_{e}\}_{e\in\mathcal{E}_{\mathrm{act}}}$.
The basic idea is that this choice of matrices $\mathbf{L}_{e}$ preserves
the signal dimensions carrying the least compression noise.

\begin{algorithm}
\caption{DPR-rank-$d_{e}$: Algorithm for the DPR scheme with rank constraints
$d_{e}<d_{\mathrm{tail}(e)}$ for some active edges $e\in\mathcal{E}_{\mathrm{act}}$}

1. Run the MM algorithm described in Algorithm 2 for fixed transformers
$\mathbf{L}_{e}=\mathbf{I}$, $e\in\mathcal{E}_{\mathrm{act}}$, without
consideration on the dimensionality constraints and denote the obtained
covariances by $\{\tilde{\mathbf{\Omega}}_{e}\}_{e\in\mathcal{E}_{\mathrm{act}}}$.

2. Fix the linear transformer $\mathbf{L}_{e}=\mathbf{V}_{e}^{\dagger}$
for $e\in\mathcal{E}_{\mathrm{act}}$ where the columns of the matrix
$\mathbf{V}_{e}$ are obtained as the eigenvectors of $\tilde{\mathbf{\Omega}}_{e}$
corresponding to the smallest $d_{e}$ eigenvalues.

3. Optimize the covariances $\{\mathbf{\Omega}_{e}\}_{e\in\mathcal{E}_{\mathrm{act}}}$
for fixed linear transform matrices $\{\mathbf{L}_{e}\}_{e\in\mathcal{E}_{\mathrm{act}}}$
using the same approach as in Step 1.
\end{algorithm}

\subsection{Multiple Control-Unit Case\label{sub:Multiple-Control-Unit-Case}}

\begin{figure}
\centering\includegraphics[width=12cm,height=9cm]{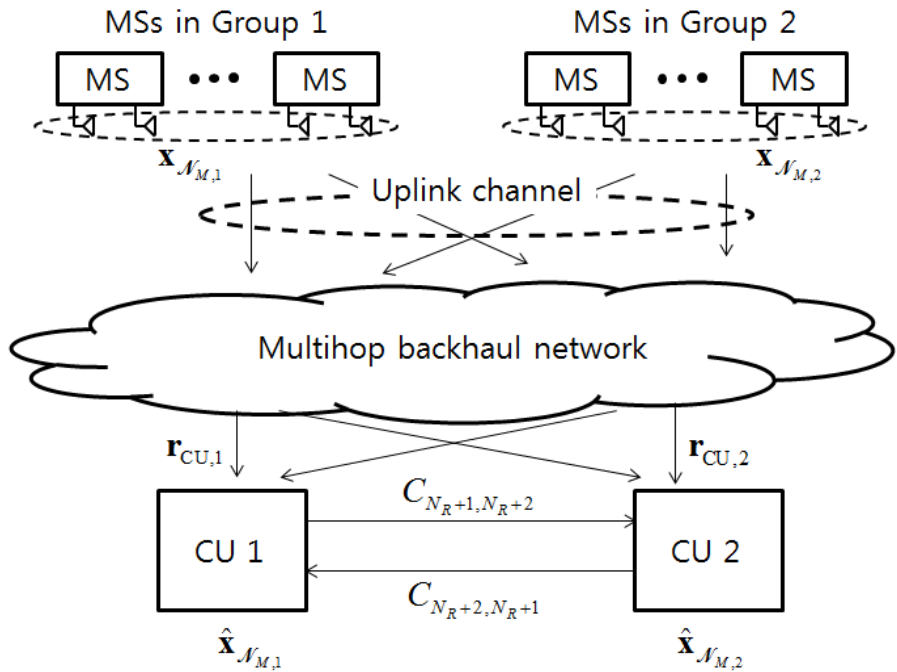}

\caption{{\footnotesize{\label{fig:system-model-multi-CU}Illustration of the
uplink of C-RANs with a multihop backhaul network and with $N_{C}=2$
CUs.}}}
\end{figure}

In this subsection, we consider the case in which there are $N_{C}$
CUs, where each CU $j$ is in charge of decoding the signals $\mathbf{x}_{\mathcal{N}_{M,j}}$
sent by a disjoint subset of MSs $\mathcal{N}_{M,j}$. We thus have
$\cup_{j=1}^{N_{C}}\mathcal{N}_{M,j}=\mathcal{N}_{M}$ and $\mathcal{N}_{M,i}\cap\mathcal{N}_{M,j}=\textrm{�}$
for all $i\neq j\in\mathcal{N}_{C}\triangleq\{1,\ldots,N_{C}\}$,
as illustrated in Fig. \ref{fig:system-model-multi-CU} for the case
with $N_{C}=2$. We assume that the CUs $i$ and $j$, with $1\leq i,j\leq N_{C}$,
which are denoted as nodes $N_{R}+i$ and $N_{R}+j$, respectively,
are connected to each other via orthogonal duplex backhaul links of
capacities $C_{N_{R}+i,N_{R}+j}$ and $C_{N_{R}+j,N_{R}+i}$ bits/s/Hz.
These links enable cooperation among the CUs for the purpose of decoding,
similar to \cite{Draper-et-al}\cite{Simeone-et-al:inter-BS}\cite{Wang-Tse}.
If the CUs perform DPR in order to communicate with one another and
we treat the MSs' messages intended for the other CUs as noise, the
problem of designing the DPR strategy at RUs and CUs can be dealt
with within the same framework studied above. The details follow easily
from the discussion above and are not provided here. Related numerical
results can be found in Sec. \ref{sec:Numerical-Results}.

\section{Decentralized Optimization\label{sec:Decentralized-Optimization}}

In this section, we discuss decentralized algorithms to address the
problem (\ref{eq:problem-DR}) for the DPR scheme studied in Sec.
\ref{sec:Decompress-and-Recompress}. For ease of notation, define
as $\mathsf{\Omega}_{i}\triangleq\{\mathbf{\Omega}_{e}\}_{e\in\Gamma_{O}(i)}$
the set of the variables describing the compression strategies at
RU $i$. In the proposed decentralized approach, the compression strategies
$\{\mathsf{\Omega}_{i}\}_{i\in\mathcal{V}}$ are determined successively
so that the variables $\{\mathsf{\Omega}_{i}\}_{i\in\mathcal{V}_{l}}$
corresponding to a layer $l$ are optimized after all variables $\{\mathsf{\Omega}_{j}\}_{j\in\cup_{m=1}^{l-1}\mathcal{V}_{m}}$
of the previous layers. In the following, we present two different
decentralized strategies that differ in the required overhead to collect
the necessary CSI from other nodes.

\begin{figure}
\centering\includegraphics[width=16cm,height=6cm]{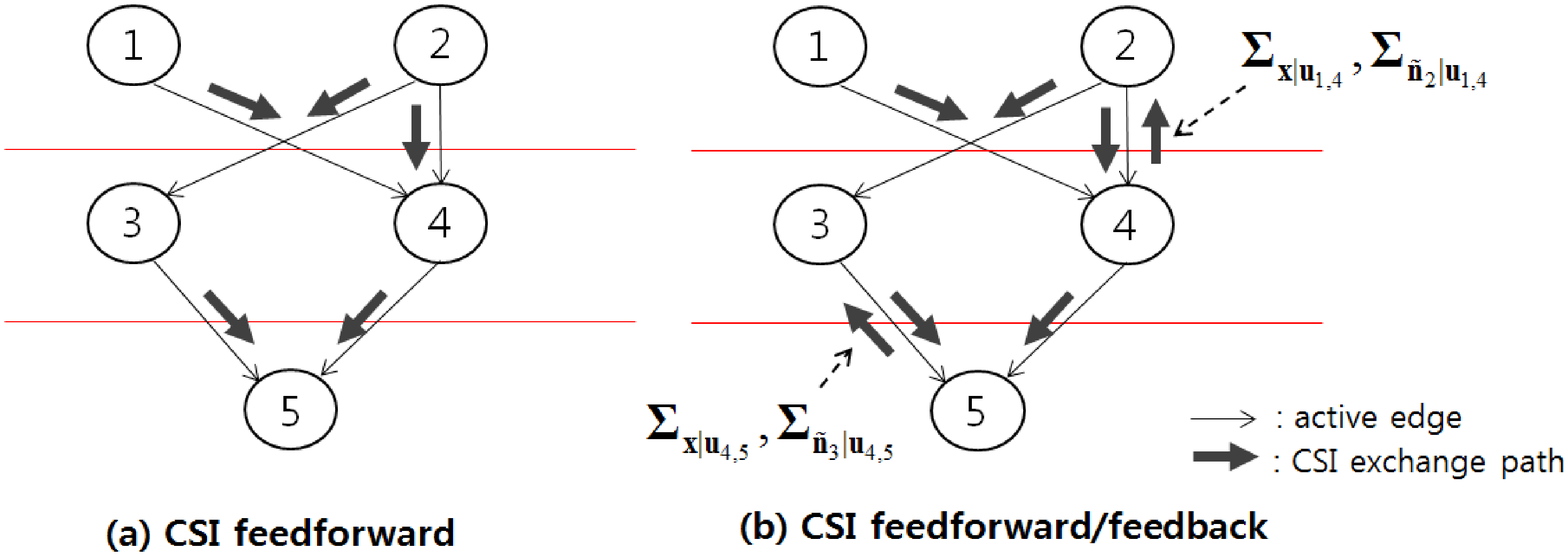}

\caption{{\footnotesize{\label{fig:CSI-FF-FB}Illustration of the CSI exchanges
required by the decentralized schemes discussed in Sec. \ref{sec:Decentralized-Optimization}
for $N_{R}=4$ RUs and routing strategy $\mathcal{V}_{1}=\{1,2\}$,
$\mathcal{V}_{2}=\{3,4\}$ and $\mathcal{V}_{3}=\{5\}$: (a) With
CSI feedforward, the CSI flows downstream from the nodes to the CU
(node 5); (b) For the case where both CSI feedforward and feedback
are allowed, nodes 2 and 3 receive additional CSI from nodes 4 and
5, respectively, under the assumption of a successive optimization
with an ordering $\pi_{1}(1)=1$, $\pi_{1}(2)=2$, $\pi_{2}(1)=4$
and $\pi_{2}(2)=3$.}}}
\end{figure}

\subsection{CSI Feedforward\label{sub:CSI-Feedforward}}

We first present a decentralized scheme in which each RU $i$ optimizes
its compression strategies $\mathsf{\Omega}_{i}$ based only on its
local CSI $\mathbf{H}_{i}$ and on the CSI \textit{fed forward} by
its ascendant nodes $\mathrm{ASC}(i)$, as illustrated in Fig. \ref{fig:CSI-FF-FB}-(a).
The set $\mathrm{ASC}(i)$ of ascendant nodes of RU $i$ is defined
as
\[
\mathrm{ASC}(i)=\left\{ i^{\prime}\in\mathcal{V}\Bigg|\begin{array}{c}
\mathrm{there\,\, exists\,\, a\,\, sequence}\,\,(i_{1},\ldots,i_{K})\in\mathcal{V}^{K}\,\,\mathrm{for\,\, some}\,\, K\\
\mathrm{such\,\, that}\,\,\left\{ (i^{\prime},i_{1}),(i_{1},i_{2}),\ldots,(i_{K-1},i_{K}),(i_{K},i)\right\} \subset\mathcal{E}_{\mathrm{act}}.
\end{array}\right\} ,
\]
and hence it includes all the nodes for which there exists an active
path to node $i$. For instance, we have $\mathrm{ASC}(4)=\{1,2\}$
in Fig. \ref{fig:CSI-FF-FB}.

Let us consider the optimization of a compression strategy $\mathbf{\Omega}_{e}$
at RU $i\in\mathcal{V}_{l}$ for an outgoing edge $e\in\Gamma_{O}(i)$.
The RUs and hence variables $\{\mathsf{\Omega}_{j}\}_{j\in\cup_{m=1}^{l-1}\mathcal{V}_{m}}$
in the previous layers have been optimized and fixed. We propose to
optimize the mutual information $I(\mathbf{x};\mathbf{u}_{e})$ at
RU $i$. This represents the sum-rate that would be achievable if
the receiving RU $\mathrm{head}(e)$ was in fact the CU and if decoding
at RU $\mathrm{head}(e)$ was based on the compressed signal $\mathbf{u}_{e}$.
This problem is stated as \begin{subequations}\label{eq:problem-decentralized-simple}
\begin{align}
\underset{\mathbf{\Omega}_{e}\succeq\mathbf{0}}{\mathrm{maximize}}\,\,\, & I(\mathbf{x};\mathbf{u}_{e})\label{eq:problem-decentralized-obj-simple}\\
\mathrm{s.t.}\,\,\,\, & I(\mathbf{r}_{i};\mathbf{u}_{e})\leq\tilde{C}_{e},\label{eq:problem-decentralized-backhaul-simple}
\end{align}
\end{subequations}where the constraint (\ref{eq:problem-decentralized-backhaul-simple})
imposes the signal $\mathbf{u}_{e}$ be compressed to a rate smaller
or equal to the backhaul capacity $\tilde{C}_{e}$. As it will be
seen, all the quantities in (\ref{eq:problem-decentralized-simple})
can be computed based on the CSI available at RU $i$. Specifically,
in order to evaluate the quantities in (\ref{eq:problem-decentralized-simple}),
we write the signal $\mathbf{r}_{i}$ in (\ref{eq:received-signal-aggregated})
to be compressed as
\begin{equation}
\mathbf{r}_{i}=\tilde{\mathbf{H}}_{i}\mathbf{x}+\tilde{\mathbf{n}}_{i},
\end{equation}
with $\tilde{\mathbf{n}}_{i}\sim\mathcal{CN}(\mathbf{0},\mathbf{\Sigma}_{\tilde{\mathbf{n}}_{i}})$,
where the effective channel matrix $\tilde{\mathbf{H}}_{i}$ and the
covariance $\mathbf{\Sigma}_{\tilde{\mathbf{n}}_{i}}$ of the effective
noise $\tilde{\mathbf{n}}_{i}$ are given as
\begin{equation}
\tilde{\mathbf{H}}_{i}=\mathbf{T}_{i}\mathbf{H}_{\mathrm{ASC}(i)}\,\,\mathrm{and}\,\,\mathbf{\Sigma}_{\tilde{\mathbf{n}}_{i}}=\mathbf{T}_{i}\mathbf{T}_{i}^{\dagger}+\tilde{\mathbf{T}}_{i}\mathrm{diag}\left(\{\mathbf{\Omega}_{e}\}_{e\in\Gamma_{O}(j),j\in\mathrm{ASC}(i)}\right)\tilde{\mathbf{T}}_{i}^{\dagger}.
\end{equation}
The matrices $\mathbf{T}_{i}$ and $\tilde{\mathbf{T}}_{i}$ are obtained
similar to (\ref{eq:transfer-matrices-DR}) by considering only the
subnetwork comprising of the ascendant nodes $\mathrm{ASC}(i)$ and
the RU $i$. Finally, the quantities appearing in the problem (\ref{eq:problem-decentralized-simple})
are obtained as
\begin{align}
I(\mathbf{x};\mathbf{u}_{e})= & \log\det\left(\tilde{\mathbf{H}}_{i}\mathbf{\Sigma}_{\mathbf{x}}\tilde{\mathbf{H}}_{i}^{\dagger}+\mathbf{\Sigma}_{\tilde{\mathbf{n}}_{i}}+\mathbf{\Omega}_{e}\right)-\log\det\left(\mathbf{\Sigma}_{\tilde{\mathbf{n}}_{i}}+\mathbf{\Omega}_{e}\right),\label{eq:equality-decentralized-simple-1}\\
\mathrm{and}\,\, I(\mathbf{r}_{i};\mathbf{u}_{e})= & \log\det\left(\tilde{\mathbf{H}}_{i}\mathbf{\Sigma}_{\mathbf{x}}\tilde{\mathbf{H}}_{i}^{\dagger}+\mathbf{\Sigma}_{\tilde{\mathbf{n}}_{i}}+\mathbf{\Omega}_{e}\right)-\log\det\left(\mathbf{\Omega}_{e}\right).\label{eq:equality-decentralized-simple-2}
\end{align}

It was shown in \cite{dCoso}\cite{Chechik-et-al} that the optimal
solution for problem (\ref{eq:problem-decentralized-simple}) is given
as
\begin{align}
\mathbf{\Omega}_{e}= & \mathbf{\Sigma}_{\tilde{\mathbf{n}}_{i}}^{1/2}\mathbf{V}_{i}\mathrm{diag}\left(\alpha_{e,1}^{-1},\ldots,\alpha_{e,d_{i}}^{-1}\right)\mathbf{V}_{i}^{\dagger}\mathbf{\Sigma}_{\tilde{\mathbf{n}}_{i}}^{1/2},\label{eq:solution-dec-FF}
\end{align}
where we defined the eigenvalue decomposition $\mathbf{\Sigma}_{\tilde{\mathbf{n}}_{i}}^{-1/2}\tilde{\mathbf{H}}_{i}\mathbf{\Sigma}_{\mathrm{x}}\tilde{\mathbf{H}}_{i}^{\dagger}\mathbf{\Sigma}_{\tilde{\mathbf{n}}_{i}}^{-1/2}+\mathbf{I}=\mathbf{V}_{i}\mathrm{diag}(\lambda_{i,1},\ldots,\lambda_{i,d_{i}})\mathbf{V}_{i}^{\dagger}$
with $\mathbf{V}_{i}\mathbf{V}_{i}^{\dagger}=\mathbf{V}_{i}^{\dagger}\mathbf{V}_{i}=\mathbf{I}$
and $\lambda_{i,1}\geq\ldots\geq\lambda_{i,d_{i}}\geq0$, and the
diagonal elements $\alpha_{e,1},\ldots,\alpha_{e,d_{i}}$ are given
as
\begin{equation}
\alpha_{e,j}=\left[\frac{1}{\mu}\left(1-\frac{1}{\lambda_{j}}\right)-1\right]^{+}
\end{equation}
for $j\in\{1,\ldots,d_{i}\}$ with $\mu$ chosen such that $\sum_{j=1}^{d_{i}}\log(1+\alpha_{e,j}\lambda_{j})=\tilde{C}_{e}$
is satisfied. The details of the decentralized algorithm proposed
in this subsection, which is referred to as ``DPR-dec-FF'', are
provided in Algorithm 4.

\begin{algorithm}
\caption{DPR-dec-FF: Decentralized algorithm with CSI feedforward}

For $l\in\{1,2,\ldots,L-1\}$,

~~For $i\in\mathcal{V}_{l}$,
\begin{itemize}
\item RU $i$ obtains the information about the matrices $\mathbf{H}_{j}$
and $\mathbf{\Omega}_{j}$ from the node $j$ for all $j\in\mathrm{ASC}(i)$.
\item RU $i$ computes the covariance $\mathbf{\Omega}_{e}$ according to
(\ref{eq:solution-dec-FF}) for $e\in\Gamma_{O}(i)$.
\end{itemize}
~~End

End
\end{algorithm}

\subsection{CSI Feedforward and Feedback\label{sub:CSI-Feedforward-and-Feedback}}

In this subsection, we discuss a decentralized approach that requires
an increased overhead for the CSI exchange as compared to the strategy
studied above. Specifically, we assume that, when optimizing the DPR
strategy for outgoing edge $e\in\Gamma_{O}(i)$, each RU $i$ in layer
$l$ is able to utilize, in addition to the local CSI $\mathbf{H}_{i}$
and the CSI $\{\mathbf{H}_{j}\}_{j\in\mathrm{ASC}(i)}$ of the ascendant
nodes, also some CSI, to be detailed below, fed back by its recipient
node $\mathrm{head}(e)$. An example of CSI exchange is illustrated
in Fig. \ref{fig:CSI-FF-FB}-(b).

To leverage the increased CSI and regulate the exchange of CSI, we
assume that the variables $\{\mathsf{\Omega}_{i}\}_{i\in\mathcal{V}_{l}}$
in each layer $l$ are successively optimized with an order $\mathsf{\Omega}_{\pi_{l}(1)}\rightarrow\ldots\rightarrow\mathsf{\Omega}_{\pi_{l}(|\mathcal{V}_{l}|)}$,
where $\pi_{l}:\{1,\ldots,|\mathcal{V}_{l}|\}\rightarrow\mathcal{V}_{l}$
denotes a permutation of the RUs in layer $\mathcal{V}_{l}$. The
idea is that the recipient node $\mathrm{head}(e)$ feeds back CSI
about the signals that have already been processed according to this
ordering. In the example in Fig. \ref{fig:CSI-FF-FB}-(b) with three
layers $\mathcal{V}_{1}=\{1,2\}$, $\mathcal{V}_{2}=\{3,4\}$ and
$\mathcal{V}_{3}=\{5\}$, we set the permutations $\pi_{1}$ and $\pi_{2}$
as $\pi_{1}(1)=1$, $\pi_{1}(2)=2$, $\pi_{2}(1)=4$ and $\pi_{2}(2)=3$
so that the compression strategies $\{\mathsf{\Omega}_{i}\}_{i\in\mathcal{N}_{R}}$
are optimized with the ordering $\mathsf{\Omega}_{1}\rightarrow\mathsf{\Omega}_{2}\rightarrow\mathsf{\Omega}_{4}\rightarrow\mathsf{\Omega}_{3}$.
We assume that the permutations $\pi_{1},\ldots,\pi_{L}$ are fixed.

Consider the optimization of the compression strategy $\mathbf{\Omega}_{e}$
for an outgoing edge $e\in\Gamma_{O}(\pi_{l}(i))$ of the $i$th RU
$\pi_{l}(i)$ in layer $l$ for given (previously optimized) variables
$\mathsf{\Omega}_{\pi_{l}(1)},\ldots,\mathsf{\Omega}_{\pi_{l}(i-1)}$
in the same layer and $\{\mathsf{\Omega}_{j}\}_{j\in\cup_{m=1}^{l-1}\mathcal{V}_{m}}$
in the previous layers. To this end, extending the approach in Sec.
\ref{sub:CSI-Feedforward}, we adopt the mutual information $I(\mathbf{x};\mathbf{u}_{e},\mathbf{v}_{e})$
as the objective function, that is, the rate that would be achieved
if RU $\mathrm{head}(e)$ was the CU decoding based on the received
signals $\mathbf{u}_{e}$ and $\mathbf{v}_{e}$. The signals $\mathbf{v}_{e}=\{\mathbf{u}_{e^{\prime}}\}_{e^{\prime}\in\mathcal{S}_{e}}$
are received by RU $\mathrm{head}(e)$ on the set $\mathcal{S}_{e}$
of all active edges to $\mathrm{head}(e)$ whose DPR strategy has
already been optimized, namely
\begin{equation}
\mathcal{S}_{e}=\left[\left(\cup_{j=1}^{i-1}\Gamma_{O}(\pi_{l}(j))\right)\cup\left(\cup_{m=1}^{l-1}\cup_{j\in\mathcal{V}_{m}}\Gamma_{O}(j)\right)\right]\cap\Gamma_{I}(\mathrm{head}(e)).\label{eq:set-already-available}
\end{equation}
The problem of optimizing $\mathbf{\Omega}_{e}$ at node $\pi_{l}(i)$
is then formulated as\begin{subequations}\label{eq:problem-decentralized-better}
\begin{align}
\underset{\mathbf{\Omega}_{e}\succeq\mathbf{0}}{\mathrm{maximize}}\,\,\, & I(\mathbf{x};\mathbf{u}_{e},\mathbf{v}_{e})\label{eq:problem-decentralized-obj-better}\\
\mathrm{s.t.}\,\,\,\, & I(\mathbf{r}_{\pi_{l}(i)};\mathbf{u}_{e})\leq\tilde{C}_{e}.\label{eq:problem-decentralized-backhaul-better}
\end{align}
\end{subequations}

The quantities in (\ref{eq:problem-decentralized-better}) can be
evaluated based on the CSI fed forward by the nodes in $\mathrm{ASC}(\pi_{l}(i))$
and on the following matrices fed back by the node $\mathrm{head}(e)$:
\begin{align}
\mathbf{\Sigma}_{\mathbf{x}|\mathbf{v}_{e}} & =\mathbf{\Sigma}_{\mathbf{x}}-\mathbf{\Sigma}_{\mathbf{x},\mathbf{v}_{e}}\mathbf{\Sigma}_{\mathbf{v}_{e}}^{-1}\mathbf{\Sigma}_{\mathbf{x},\mathbf{v}_{e}}^{\dagger},\label{eq:conditional-covariance-1}\\
\mathrm{and}\,\,\mathbf{\Sigma}_{\tilde{\mathbf{n}}_{\pi_{l}(i)}|\mathbf{v}_{e}} & =\mathbf{\Sigma}_{\tilde{\mathbf{n}}_{\pi_{l}(i)}}-\mathbf{\Sigma}_{\tilde{\mathbf{n}}_{\pi_{l}(i)},\mathbf{v}_{e}}\mathbf{\Sigma}_{\mathbf{v}_{e}}^{-1}\mathbf{\Sigma}_{\tilde{\mathbf{n}}_{\pi_{l}(i)}|\mathbf{v}_{e}}^{\dagger},\label{eq:conditional-covariance-2}
\end{align}
where the detailed computation of the matrices $\mathbf{\Sigma}_{\mathbf{x},\mathbf{v}_{e}}$,
$\mathbf{\Sigma}_{\tilde{\mathbf{n}}_{\pi_{l}(i)},\mathbf{v}_{e}}$
and $\mathbf{\Sigma}_{\mathbf{v}_{e}}$ is presented in Appendix \ref{appendix:calculation-correlation}.
In fact, using the chain rule of mutual information, we can decompose
the objective function as $I(\mathbf{x};\mathbf{u}_{e},\mathbf{v}_{e})=I(\mathbf{x};\mathbf{u}_{e}|\mathbf{v}_{e})+I(\mathbf{x};\mathbf{v}_{e})$.
Since the second term of the right-hand side does not depend on $\mathbf{\Omega}_{e}$,
we can replace the objective with $I(\mathbf{x};\mathbf{u}_{e}|\mathbf{v}_{e})$,
which is calculated as
\begin{align}
f_{e}^{\mathrm{dec}}(\mathbf{\Omega}_{e}) & \triangleq I(\mathbf{x};\mathbf{u}_{e}|\mathbf{v}_{e})\label{eq:equality-decentralized-better}\\
 & =\log\det\left(\tilde{\mathbf{H}}_{\pi_{l}(i)}\mathbf{\Sigma}_{\mathbf{x}|\mathbf{v}_{e}}\tilde{\mathbf{H}}_{\pi_{l}(i)}^{\dagger}+\mathbf{\Sigma}_{\tilde{\mathbf{n}}_{\pi_{l}(i)}|\mathbf{v}_{e}}+\mathbf{\Omega}_{e}\right)-\log\det\left(\mathbf{\Sigma}_{\tilde{\mathbf{n}}_{\pi_{l}(i)}|\mathbf{v}_{e}}+\mathbf{\Omega}_{e}\right).\nonumber
\end{align}
Also, the left-hand side $I(\mathbf{r}_{\pi_{l}(i)};\mathbf{u}_{e})$
of (\ref{eq:problem-decentralized-backhaul-better}) can be calculated
as (\ref{eq:equality-decentralized-simple-2}) with the index $i$
replaced with $\pi_{l}(i)$. Substituting (\ref{eq:equality-decentralized-simple-2})
and (\ref{eq:equality-decentralized-better}) into (\ref{eq:problem-decentralized-better})
leads to a difference-of-convex problem, and thus we can use the MM
approach \cite{Beck} to find a stationary point of the problem. The
algorithm for the decentralized scheme proposed in this subsection,
which is referred to as ``DPR-dec-FF-FB'', is presented in Algorithm
5, where we define the functions $\tilde{f}_{e}^{\mathrm{dec}}(\mathbf{\Omega}_{e}^{(t+1)},\mathbf{\Omega}_{e}^{(t)})$
and $\tilde{g}_{e}^{\mathrm{dec}}(\mathbf{\Omega}_{e}^{(t+1)},\mathbf{\Omega}_{e}^{(t)})$
as
\begin{align}
\tilde{f}_{e}^{\mathrm{dec}}(\mathbf{\Omega}_{e}^{(t+1)},\mathbf{\Omega}_{e}^{(t)}) & =\log\det\left(\tilde{\mathbf{H}}_{\pi_{l}(i)}\mathbf{\Sigma}_{\mathbf{x}|\mathbf{v}_{e}}\tilde{\mathbf{H}}_{\pi_{l}(i)}^{\dagger}+\mathbf{\Sigma}_{\tilde{\mathbf{n}}_{\pi_{l}(i)}|\mathbf{v}_{e}}+\mathbf{\Omega}_{e}^{(t+1)}\right)\\
 & -\varphi\left(\mathbf{\Sigma}_{\tilde{\mathbf{n}}_{\pi_{l}(i)}|\mathbf{v}_{e}}+\mathbf{\Omega}_{e}^{(t+1)},\mathbf{\Sigma}_{\tilde{\mathbf{n}}_{\pi_{l}(i)}|\mathbf{v}_{e}}+\mathbf{\Omega}_{e}^{(t)}\right),\nonumber \\
\mathrm{and}\,\,\tilde{g}_{e}^{\mathrm{dec}}(\mathbf{\Omega}_{e}^{(t+1)},\mathbf{\Omega}_{e}^{(t)}) & =\varphi\left(\tilde{\mathbf{H}}_{\pi_{l}(i)}\mathbf{\Sigma}_{\mathbf{x}}\tilde{\mathbf{H}}_{\pi_{l}(i)}^{\dagger}+\mathbf{\Sigma}_{\tilde{\mathbf{n}}_{\pi_{l}(i)}}+\mathbf{\Omega}_{e}^{(t+1)},\tilde{\mathbf{H}}_{\pi_{l}(i)}\mathbf{\Sigma}_{\mathbf{x}}\tilde{\mathbf{H}}_{\pi_{l}(i)}^{\dagger}+\mathbf{\Sigma}_{\tilde{\mathbf{n}}_{\pi_{l}(i)}}+\mathbf{\Omega}_{e}^{(t)}\right)\nonumber \\
 & -\log\det\left(\mathbf{\Omega}_{e}^{(t+1)}\right).
\end{align}

\begin{algorithm}
\caption{DPR-dec-FF-FB: Decentralized algorithm with both CSI feedforward and
feedback}

For $l\in\{1,2,\ldots,L-1\}$,

~~For $i\in\{1,2,\ldots,|\mathcal{V}_{l}|\}$,
\begin{itemize}
\item RU $\pi_{l}(i)$ obtains the information about the matrices $\mathbf{H}_{j}$
and $\mathbf{\Omega}_{j}$ from the node $j$ for all $j\in\mathrm{ASC}(\pi_{l}(i))$.
\item RU $\pi_{l}(i)$ obtains the information about the matrices $\mathbf{\Sigma}_{\mathbf{x}|\mathbf{v}_{e}}$
and $\mathbf{\Sigma}_{\tilde{\mathbf{n}}_{\pi_{l}(i)}|\mathbf{v}_{e}}$
from the node $\mathrm{head}(e)$ for all outgoing edges $e\in\Gamma_{O}(\pi_{l}(i))$.
\item For all $e\in\Gamma_{O}(\pi_{l}(i))$, RU $\pi_{l}(i)$ updates the
covariance $\mathbf{\Omega}_{e}$ according to the following MM algorithm
for problem (\ref{eq:problem-decentralized-better}):

\begin{itemize}
\item 1. Initialize the matrix $\mathbf{\Omega}_{e}^{(1)}$ to an arbitrary
feasible positive semidefinite matrix for problem (\ref{eq:problem-decentralized-better})
and set $t=1$.
\item 2. Update the matrices $\mathbf{\Omega}_{e}^{(t+1)}$ as a solution
of the following (convex) problem
\begin{align}
\underset{\mathbf{\Omega}_{e}^{(t+1)}\succeq\mathbf{0}}{\mathrm{maximize}}\,\,\, & \tilde{f}_{e}^{\mathrm{dec}}(\mathbf{\Omega}_{e}^{(t+1)},\mathbf{\Omega}_{e}^{(t)})\label{eq:problem-convexified-centralized-2}\\
\mathrm{s.t.}\,\,\,\,\,\,\,\,\, & \tilde{g}_{e}^{\mathrm{dec}}(\mathbf{\Omega}_{e}^{(t+1)},\mathbf{\Omega}_{e}^{(t)})\leq\tilde{C}_{e}.\nonumber
\end{align}

\item 3. Stop if a convergence criterion is satisfied. Otherwise, set $t\leftarrow t+1$
and go back to Step 2.
\end{itemize}
\end{itemize}
~~End

End
\end{algorithm}

\subsection{Utilizing Side Information for Decompression\label{sub:Side-Information}}

In this subsection, we investigate the performance advantage of compression
with side information. As discussed in \cite{Sanderovich}\cite{dCoso}
for backhaul networks with a star topology, the use of side information
for decompression via Wyner-Ziv coding/decoding \cite[Ch. 12]{ElGamal-Kim}
improves the efficiency of the backhaul link utilization by leveraging
the correlation of the received baseband signals at the RUs. Similar
to \cite{WeiYu}\cite{Chen-Berger}, we assume that each RU $i$ successively
recovers the incoming compressed signals $\{\mathbf{u}_{e}\}_{e\in\Gamma_{I}(i)}$
with an order $\tilde{\pi}_{i}:\{1,\ldots,|\Gamma_{I}(i)|\}\rightarrow\Gamma_{I}(i)$
(i.e., $\mathbf{u}_{\tilde{\pi}_{i}(1)}\rightarrow\ldots\rightarrow\mathbf{u}_{\tilde{\pi}_{i}(|\Gamma_{I}(i)|)}$).
Using this order, as in the rest of this section, a successive optimization
approach is adopted whereby each variable $\mathbf{\Omega}_{\tilde{\pi}_{i}(j)}$
corresponding to an edge $\tilde{\pi}_{i}(j)$ is optimized at RU
$\mathrm{tail}(\tilde{\pi}_{i}(j))$ after the variables $\mathbf{\Omega}_{\tilde{\pi}_{i}(1)},\ldots,\mathbf{\Omega}_{\tilde{\pi}_{i}(j-1)}$
and $\{\mathsf{\Omega}_{j}\}_{j\in\cup_{m=1}^{l-1}\mathcal{V}_{m}}$.

Let us consider the optimization of the compression covariance $\mathbf{\Omega}_{e}$
for an outgoing edge $e\in\Gamma_{O}(i)$ at RU $i$. We define as
$\mathbf{v}_{e}=\{\mathbf{u}_{e^{\prime}}\}_{e^{\prime}\in\tilde{\mathcal{S}}_{e}}$
the signals available at the receiving node $\mathrm{head}(e)$ when
decompressing the signal $\mathbf{u}_{e}$, where we have $\tilde{\mathcal{S}}_{e}=\{\tilde{\pi}_{\mathrm{head}(e)}(1),\ldots,\tilde{\pi}_{\mathrm{head}(e)}(\tilde{\pi}_{\mathrm{head}(e)}^{-1}(e)-1)\}$.
As for the discussion in Sec. \ref{sub:CSI-Feedforward-and-Feedback},
we aim at maximizing the mutual information $I(\mathbf{x};\mathbf{u}_{e},\mathbf{v}_{e})$,
which measures the sum-rate achievable under the assumption that the
RU $\mathrm{head}(e)$ is the CU and that it performs decoding of
the MSs' signals $\mathbf{x}$ based on the signals $\mathbf{u}_{e}$
and $\mathbf{v}_{e}$. Then, the problem of optimizing $\mathbf{\Omega}_{e}$
at RU $i$ is stated as (\ref{eq:problem-decentralized-better}) with
the constraint (\ref{eq:problem-decentralized-backhaul-better}) replaced
by the condition
\begin{align}
 & I(\mathbf{r}_{i};\mathbf{u}_{e}|\mathbf{v}_{e})\leq\tilde{C}_{e}.\label{eq:constraint-SI}
\end{align}
By the Wyner-Ziv theorem \cite[Ch. 12]{ElGamal-Kim}, this constraint
guarantees that the signal $\mathbf{u}_{e}$ can be successfully recovered
by RU $\mathrm{head}(e)$ if the latter utilizes the signal $\mathbf{v}_{e}$
as side information when decompressing. It can be shown that the constraint
(\ref{eq:constraint-SI}) can be evaluated as
\begin{align}
g_{e}^{\mathrm{dec-SI}}(\mathbf{\Omega}_{e}) & \triangleq\log\det\left(\tilde{\mathbf{H}}_{i}\mathbf{\Sigma}_{\mathbf{x}|\mathbf{v}_{e}}\tilde{\mathbf{H}}_{i}^{\dagger}+\mathbf{\Sigma}_{\tilde{\mathbf{n}}_{i}|\mathbf{v}_{e}}+\mathbf{\Omega}_{e}\right)\\
 & -\log\det\left(\mathbf{\Omega}_{e}\right)\leq\tilde{C}_{e}.\nonumber
\end{align}
The problem at hand is again a difference-of-convex problem, and hence
a stationary point of the problem can be found by following a procedure
similar to Sec. \ref{sub:CSI-Feedforward-and-Feedback}.

The algorithm for the decentralized scheme discussed in this subsection,
which is referred to as ``DPR-dec-SI'', is described in Algorithm
6, where we have defined the function $\tilde{g}_{e}^{\mathrm{dec-SI}}(\mathbf{\Omega}_{e}^{(t+1)},\mathbf{\Omega}_{e}^{(t)})$
as
\begin{align}
 & \tilde{g}_{e}^{\mathrm{dec-SI}}(\mathbf{\Omega}_{e}^{(t+1)},\mathbf{\Omega}_{e}^{(t)})\label{eq:constraint-backhaul-SI}\\
= & \varphi\left(\tilde{\mathbf{H}}_{\pi_{l}(i)}\mathbf{\Sigma}_{\mathbf{x}|\mathbf{v}_{e}}\tilde{\mathbf{H}}_{\pi_{l}(i)}^{\dagger}+\mathbf{\Sigma}_{\tilde{\mathbf{n}}_{\pi_{l}(i)}|\mathbf{v}_{e}}+\mathbf{\Omega}_{e}^{(t+1)},\tilde{\mathbf{H}}_{\pi_{l}(i)}\mathbf{\Sigma}_{\mathbf{x}|\mathbf{v}_{e}}\tilde{\mathbf{H}}_{\pi_{l}(i)}^{\dagger}+\mathbf{\Sigma}_{\tilde{\mathbf{n}}_{\pi_{l}(i)}|\mathbf{v}_{e}}+\mathbf{\Omega}_{e}^{(t)}\right)\nonumber \\
 & -\log\det\left(\mathbf{\Omega}_{e}^{(t+1)}\right).\nonumber
\end{align}
Note that the DPR scheme discussed in this subsection is equivalent
to the decentralized scheme studied in Sec. \ref{sub:CSI-Feedforward-and-Feedback}
in terms of the overhead to collect the necessary CSI from other nodes.

\begin{algorithm}
\caption{DPR-dec-SI: Decentralized algorithm that leverages side information
for decompression}

For $l\in\{1,2,\ldots,L-1\}$,

~~For $i\in\{1,2,\ldots,|\mathcal{V}_{l}|\}$,
\begin{itemize}
\item RU $\pi_{l}(i)$ obtains the information about the matrices $\mathbf{H}_{j}$
and $\mathbf{\Omega}_{j}$ from the node $j$ for all $j\in\mathrm{ASC}(\pi_{l}(i))$.
\item RU $\pi_{l}(i)$ obtains the information about the matrices $\mathbf{\Sigma}_{\mathbf{x}|\mathbf{v}_{e}}$
and $\mathbf{\Sigma}_{\tilde{\mathbf{n}}_{\pi_{l}(i)}|\mathbf{v}_{e}}$
from the node $\mathrm{head}(e)$ for all outgoing edges $e\in\Gamma_{O}(\pi_{l}(i))$.
\item For all $e\in\Gamma_{O}(\pi_{l}(i))$, RU $\pi_{l}(i)$ updates the
covariance $\mathbf{\Omega}_{e}$ according to the following MM algorithm
for problem (\ref{eq:problem-decentralized-better}) with the constraint
(\ref{eq:problem-decentralized-backhaul-better}) replaced with (\ref{eq:constraint-backhaul-SI}):

\begin{itemize}
\item 1. Initialize the matrix $\mathbf{\Omega}_{e}^{(1)}$ to an arbitrary
feasible positive semidefinite matrix for problem (\ref{eq:problem-decentralized-better})
and set $t=1$.
\item 2. Update the matrices $\mathbf{\Omega}_{e}^{(t+1)}$ as a solution
of the following (convex) problem
\begin{align}
\underset{\mathbf{\Omega}_{e}^{(t+1)}\succeq\mathbf{0}}{\mathrm{maximize}}\,\, & \tilde{f}_{e}^{\mathrm{dec}}(\mathbf{\Omega}_{e}^{(t+1)},\mathbf{\Omega}_{e}^{(t)})\label{eq:problem-convexified-centralized-2-1}\\
\mathrm{s.t.}\,\,\,\,\,\,\,\,\, & \tilde{g}_{e}^{\mathrm{dec}\mathrm{-SI}}(\mathbf{\Omega}_{e}^{(t+1)},\mathbf{\Omega}_{e}^{(t)})\leq\tilde{C}_{e}.\nonumber
\end{align}

\item 3. Stop if a convergence criterion is satisfied. Otherwise, set $t\leftarrow t+1$
and go back to Step 2.
\end{itemize}
\end{itemize}
~~End

End
\end{algorithm}

\section{Numerical Results\label{sec:Numerical-Results}}

\begin{figure}
\centering\includegraphics[width=16cm,height=8.5cm]{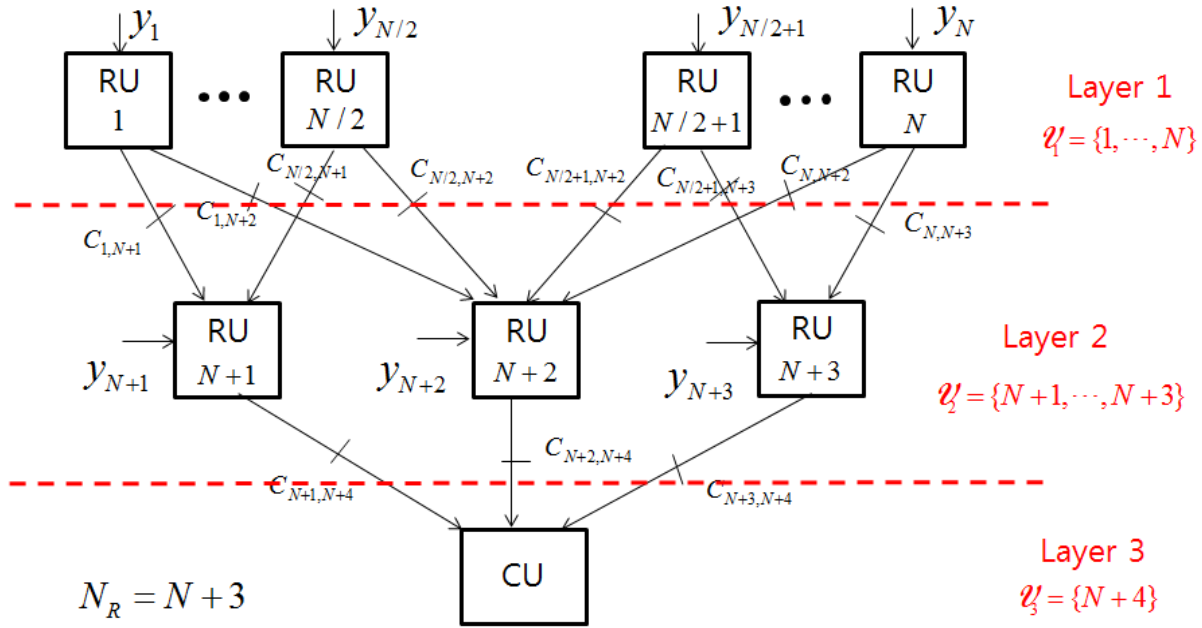}

\caption{{\footnotesize{\label{fig:example-network-simul}The hierarchical
backhaul network assumed for the simulations in Sec. \ref{sec:Numerical-Results}.
All RUs have the same received SNR and are equipped with a single
receive antenna.}}}
\end{figure}

In this section, we demonstrate the performance of the backhaul communication
schemes studied in the paper. Unless stated otherwise, we consider
the backhaul network shown in Fig. \ref{fig:example-network-simul}
with a routing strategy described by the partition $\mathcal{V}_{1}=\{1,\ldots,N\}$,
$\mathcal{V}_{2}=\{N+1,N+2,N+3\}$ and $\mathcal{V}_{3}=\{N+4\}$
that leads to all edges being activated, i.e., $\mathcal{E}=\mathcal{E}_{\mathrm{act}}$.
This scenario captures a hierarchical backhaul network in which some
RUs have direct backhaul links to the CU, i.e., the layer-2 nodes,
while the other RUs, i.e., the layer-1 nodes, are distributed over
the geographical area and connected only to the closest layer-2 nodes.
We assume that all edges have the same backhaul capacity unless stated
otherwise and set $T=D$ so that the effective capacity satisfies
the equality $\tilde{C}_{e}=C_{e}$. It is also assumed that the elements
of the channel matrix $\mathbf{H}_{i}$ are independent and identically
distributed (i.i.d.) $\mathcal{CN}(0,1)$ variables for $i\in\mathcal{N}_{R}$
(Rayleigh fading). MSs and RUs are equipped with a single antenna
and the signals $\mathbf{x}$ transmitted by MSs are distributed as
$\mathbf{x}\sim\mathcal{CN}(\mathbf{0},P_{\mathrm{tx}}\mathbf{I})$,
so that the transmitted power by each MS is given by $P_{\mathrm{tx}}$.
We focus on the average sum-rate performance measured by averaging
the instantaneous sum-rates over many channel realizations.

\begin{figure}
\centering\includegraphics[width=12cm,height=9cm]{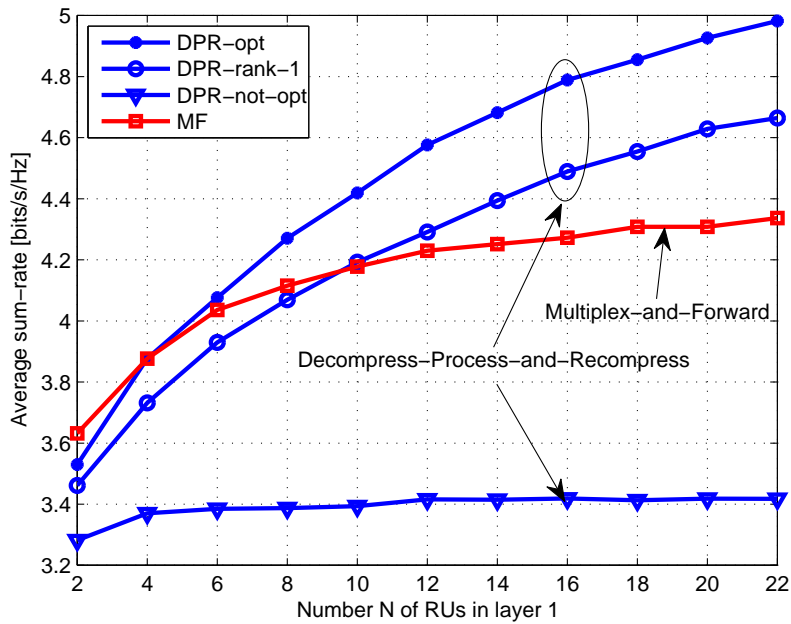}

\caption{\label{fig:graph-as-M}Average sum-rate versus the number $N$ RUs
in layer 1 with $N_{M}=4$ MSs, $P_{\mathrm{tx}}=0$ dB, $C=3$ bits/s/Hz
and RU $2N+2$ deactivated.}
\end{figure}

Fig. \ref{fig:graph-as-M} shows the average sum-rate versus the number
$N$ of RUs in layer 1 with $N_{M}=4$ MSs, $P_{\mathrm{tx}}=0$ dB
and backhaul capacity $C_{e}=3$ bits/s/Hz except for RU $N+2$ which
is assumed to be deactivated, i.e., $C_{N+2,N+4}=0$. We compare the
DPR scheme studied in Sec. \ref{sec:Decompress-and-Recompress} with
the MF scheme analyzed in Sec. \ref{sec:Forward-without-Decompress}.
For the DPR scheme, we observe the performance with the compression
strategies $\mathbf{\Omega}_{e}$ for all edges $e\in\mathcal{E}_{\mathrm{act}}$
optimized according to Algorithm 2 (labeled as ``DPR-opt''), limited-rank
processing described in Algorithm 3 with $d_{e}=1$ (labeled as ``DPR-rank-1'')
and with the compression covariances constrained to be equal to scaled
identities, i.e., $\{\mathbf{\Omega}_{e}=c_{e}\mathbf{I}\}_{e\in\mathcal{E}_{\mathrm{act}}}$
(labeled as ``DPR-not-opt''). It is first observed that the performance
gain of the DPR scheme over MF becomes more pronounced as the number
$N$ of RUs in the first layer increases. This implies that, as the
density of the RUs' deployment increases, it is desirable for each
RU in layer 2 to perform in-network processing of the signals received
from layer 1 in order to use the backhaul links to the CU more efficiently.
In a similar vein, the performance loss of DPR-not-opt scheme becomes
more significant for large $N$ since a proper allocation of compression
rates is more important in the presence of a large number of signals
sharing the backhaul capacity. We also note that the performance loss
of DPR-rank-1 compared to that of DPR-opt is relatively small even
for large $N$. As further discussed below, this is due to the fact
that the backhaul capacity $C$ is small and hence rank reduction
is effectively implemented also by DPR-opt (by setting some of the
quantization noise signals in the covariance matrices $\mathbf{\Omega}_{e}$
to be very large).

\begin{figure}
\centering\includegraphics[width=12cm,height=9cm]{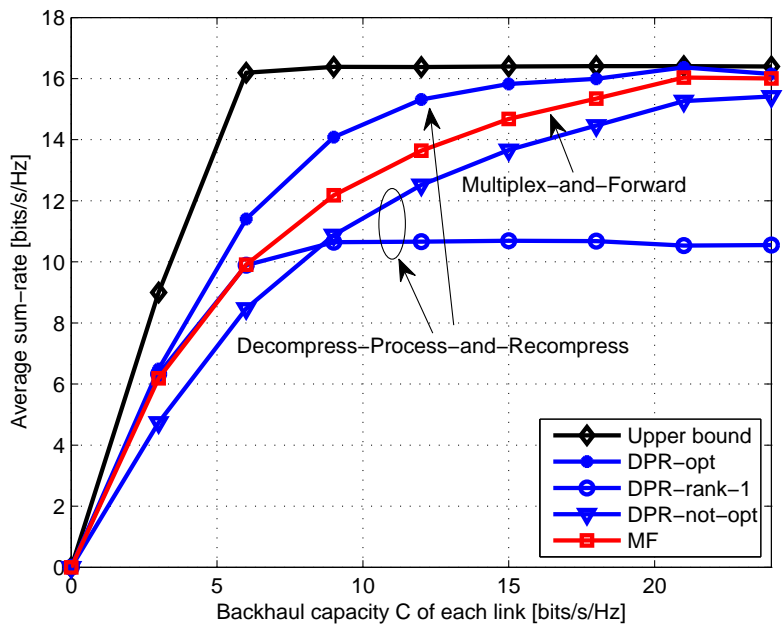}

\caption{\label{fig:graph-as-C}Average sum-rate versus the backhaul capacity
$C$ of each link with $N_{M}=5$ MSs, $N=8$ RUs in layer 1 and $P_{\mathrm{tx}}=0$
dB.}
\end{figure}

In Fig. \ref{fig:graph-as-C}, we plot the average sum-rate versus
the backhaul capacity $C$ of all edges with $N_{M}=5$ MSs, $N=8$
RUs in the first layer and $P_{\mathrm{tx}}=0$ dB. For reference,
we also plot an upper bound $R_{\mathrm{UB}}$ on the sum-rate achievable
with Gaussian quantization noises and without leveraging side information
(see Sec. \ref{sub:Side-Information}). Using cut-set arguments \cite[Theorem 14.10.1]{Cover},
this is obtained as $R_{\mathrm{UB}}=\min(\sum_{i\in\mathcal{V}_{2}}C_{i,N+4},\, R_{\mathrm{direct}})$,
where the first term is the capacity of cut-set between the RUs in
layer 2 and the CU, and the rate $R_{\mathrm{direct}}$ is computed
by assuming that every RU $i$ with $i\in\mathcal{N}_{R}$ is directly
connected to the CU $N_{R}+1$ via a backhaul link of capacity $\sum_{e\in\Gamma_{O}(i)}C_{e}$.
We first observe from Fig. \ref{fig:graph-as-C} that DPR-opt outperforms
the MF scheme in the regime of intermediate backhaul capacities $C$,
while, when the backhaul capacity $C$ is either very small or very
large, MF is sufficient. It is also seen that both DPR-opt and MF
achieve the upper bound if the backhaul capacity $C$ is large enough.
Finally, following the discussion above, we observe that, when the
backhaul capacity is sufficiently large, limiting the rank of the
baseband signals sent on the backhaul links (DPR-rank-1) leads to
a significant performance loss.

\begin{figure}
\centering\includegraphics[width=12cm,height=9cm]{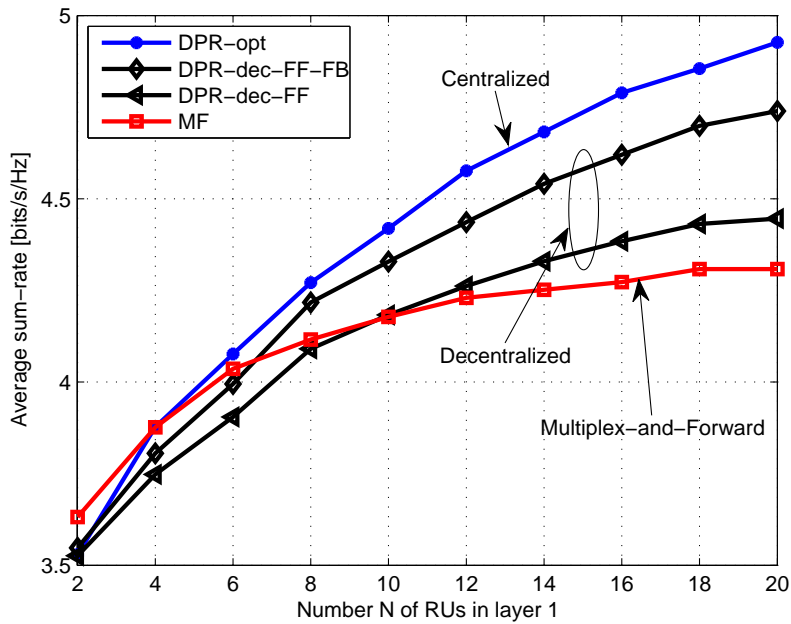}

\caption{\label{fig:graph-as-M-decentralized}Average sum-rate versus the number
$N$ of RUs in layer 1 for centralized and decentralized schemes with
$N_{M}=4$ MSs, $P_{\mathrm{tx}}=0$ dB, $C=3$ bits/s/Hz and RU $N+2$
deactivated.}
\end{figure}
We now turn to the evaluation of the performance of the decentralized
schemes studied in Sec. \ref{sec:Decentralized-Optimization}. Specifically,
in Fig. \ref{fig:graph-as-M-decentralized}, we compare the sum-rates
of DPR-opt and DPR-not-opt, the decentralized algorithm with only
CSI feedforward in Algorithm 4, labeled as DPR-dec-FF, and the decentralized
algorithm with both CSI feedforward and feedback in Algorithm 5, labeled
as DPR-dec-FF-FB. The sum-rate is shown versus the number $N$ of
RUs in layer 1 with $N_{M}=4$ MSs, $P_{\mathrm{tx}}=0$ dB, $C=3$
bits/s/Hz and RU $N+2$ deactivated. The performance loss of the decentralized
strategies becomes more pronounced as the number $N$ of RUs in the
first layer increases while still outperforming the baseline MF scheme.
Moreover, it is seen that the feedback CSI information brings significant
benefits as compared to using only feedforward CSI information.

\begin{figure}
\centering\includegraphics[width=12cm,height=9cm]{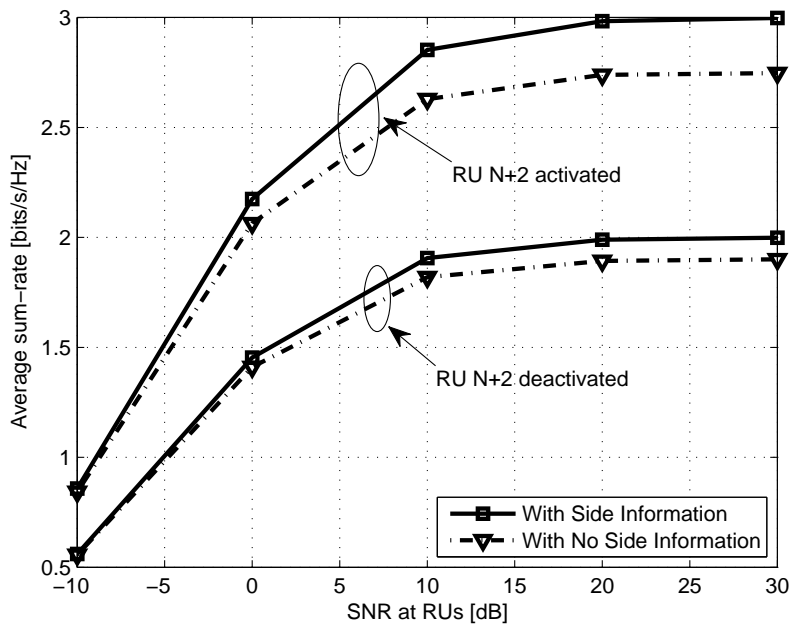}

\caption{\label{fig:graph-as-SNR}Average sum-rate achieved by the decentralized
DPR schemes, DPR-dec-FF-FB with and without side information, versus
the transmitted power $P_{\mathrm{tx}}$ by each MS with $N_{M}=4$
MSs, $N=6$ RUs in layer 1 and $C=1$ bits/s/Hz.}
\end{figure}

We now examine the advantage of utilizing side information via Wyner-Ziv
coding following the analysis in Sec. \ref{sub:Side-Information}.
Specifically, in Fig. \ref{fig:graph-as-SNR}, we plot the average
sum-rate versus the transmitted power $P_{\mathrm{tx}}$ by each MS
with $N_{M}=4$ MSs, $N=6$ RUs in the first layer and $C=1$ bits/s/Hz.
We compare the performance of DPR-dec-FF-FB discussed above with the
analogous scheme proposed in Sec. \ref{sub:Side-Information} and
described in Algorithm 6 that leverages side information for decompression.
We emphasize that, while both schemes require an equivalent overhead
for CSI exchange, only the latter scheme utilizes the side information
for decompression via Wyner-Ziv coding/decoding. From the figure,
it is seen that utilizing side information for decompression is beneficial
especially in the high SNR regime, since at low SNR, the performance
is dominated by the additive noise and the quantization noise plays
a secondary role. Moreover, this effect is more pronounced when all
RUs in layer 2 are activated due to the increased number of available
side information signals.

\begin{figure}
\centering\includegraphics[width=15cm,height=8.5cm]{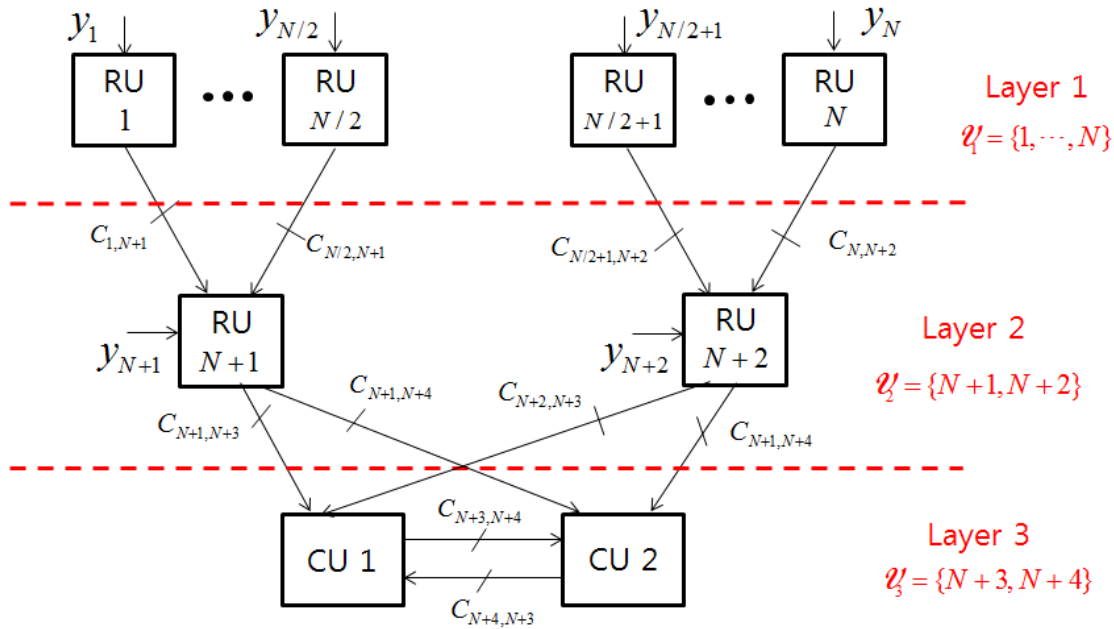}

\caption{{\footnotesize{\label{fig:example-network-simul-mulCU}The backhaul
network assumed for the simulations on the case with two CUs studied
in Sec. \ref{sub:Multiple-Control-Unit-Case}. All RUs have the same
received SNR and are equipped with a single receive antenna.}}}
\end{figure}

\begin{figure}
\centering\includegraphics[width=12cm,height=9cm]{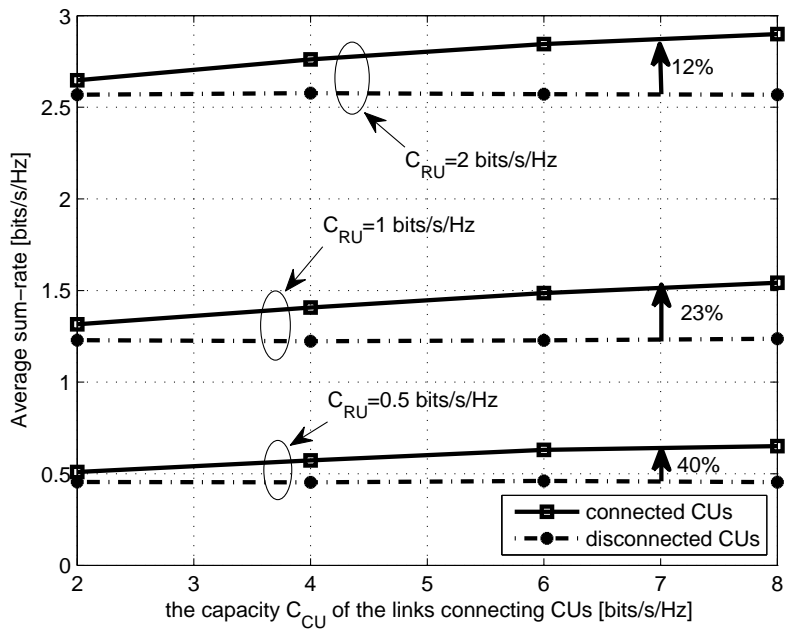}

\caption{\label{fig:graph-as-C-mulCU}Average sum-rate for the scenario in
Fig. \ref{fig:example-network-simul-mulCU} with two CUs versus the
backhaul capacity $C_{\mathrm{CU}}$ of the backhaul links between
the CUs with $N_{M,1}=N_{M,2}=2$, $N=2$ and $P_{\mathrm{tx}}=0$
dB.}
\end{figure}

Finally, in Fig. \ref{fig:graph-as-C-mulCU}, we observe the average
sum-rate performance of the DPR scheme studied in Sec. \ref{sub:Multiple-Control-Unit-Case}
for the case with multiple CUs. Specifically, we assume the backhaul
network shown in Fig. \ref{fig:example-network-simul-mulCU} in which
two CUs are connected to a common set of RUs in layer 2. Under the
assumption that all the backhaul links have the same capacity $C_{\mathrm{RU}}$
bits/s/Hz except for the backhaul links $\{(N_{R}+1,N_{R}+2),(N_{R}+2,N_{R}+1)\}$
connecting the CUs, we plot the average sum-rates versus the capacity
$C_{\mathrm{CU}}$ of the backhaul links between the CUs with $N_{M,1}=N_{M,2}=2$,
$N=2$ and $P_{\mathrm{tx}}=0$ dB. For comparison, we also plot the
sum-rate with $C_{\mathrm{CU}}=0$. It is observed that enabling cooperation
among the CUs leads to significant gains. For instance, with backhaul
capacity $C_{\mathrm{CU}}=7$ bits/s/Hz, we obtain sum-rate gains
of 40\%, 23\% and 12\% for the backhaul capacities $C_{\mathrm{RU}}=0.5$,
$1$ and $2$ bits/s/Hz, respectively. This shows that inter-CU cooperation
is able to partly compensate for a smaller backhaul capacity of the
other backhaul links.

\section{Conclusion\label{sec:Conclusion}}

In this work, we have studied efficient compression and routing strategies
for the backhaul of uplink C-RAN systems with a multihop backhaul
topology. We have first presented a baseline backhaul scheme in which
each RU forwards the bit streams received from the connected RUs without
any processing. Since this strategy may suffer from a significant
performance degradation when the backhaul network is well connected,
we have introduced a scheme in which each RU decompresses the received
bit streams and performs linear in-network processing of the decompressed
signals. To design the discussed backhaul schemes, we tackled the
sum-rate maximization problems under backhaul capacity constraints.
While the basic solutions require full CSI, decentralized optimization
algorithms were also proposed under the assumption that each RU has
limited CSI. Also, scenarios in which multiple CUs are in charge of
decoding disjoint subsets of MSs' messages were briefly dealt with.
We finally provided numerical results assessing the performance of
the considered compression schemes and specifically lending evidence
to the advantages of in-network processing scheme in the presence
of a dense deployment of RUs. We remark that it would be an important
work to study multihop backhaul compression assuming that each RU
has imperfect CSI of the other RUs or imperfect information about
the number or the capacity of outgoing backhaul links. It is expected
that, in those cases, allowing each RU to send multiple successive
refinement layers to the next nodes could be advantageous as compared
to sending a single description (see, e.g., \cite{Simeone-et-al}).

\appendices

\section{Proof of Proposition \ref{prop:identity}}\label{appendix:proof-lem-identity}

In this appendix, we show that, for any feasible variables $\{\mathbf{L}_{e}^{\prime},\mathbf{\Omega}_{e}^{\prime}\}_{e\in\mathcal{E}_{\mathrm{act}}}$,
i.e., satisfying the constraints (\ref{eq:problem-backhaul}), it
is always possible to find feasible variables $\{\mathbf{I},\mathbf{\Omega}_{e}^{\prime\prime}\}_{e\in\mathcal{E}_{\mathrm{act}}}$
that achieve the same sum-rate. To this end, we start by assuming
that the matrices $\mathbf{L}_{e}^{\prime}$ are full rank. Under
this assumption, we set the matrices $\mathbf{\Omega}_{e}^{\prime\prime}$
as
\begin{equation}
\mathbf{\Omega}_{e}^{\prime\prime}=\mathbf{G}_{e}\mathbf{\Omega}_{e}^{\prime}\mathbf{G}_{e}^{\dagger}\label{eq:Omega-double-prime}
\end{equation}
for $e\in\mathcal{E}_{\mathrm{act}}$ with the matrix $\mathbf{G}_{e}$
defined as
\begin{equation}
\mathbf{G}_{e}=\begin{cases}
\mathrm{diag}\left(\{\mathbf{G}_{\tilde{e}}\}_{\tilde{e}\in\Gamma_{I}(\mathrm{tail}(e))}\right)(\mathbf{L}_{e}^{\prime})^{-1}, & \mathrm{if}\,\,\Gamma_{i}(\mathrm{tail}(e))\neq\varnothing\\
(\mathbf{L}_{e}^{\prime})^{-1}, & \mathrm{otherwise}
\end{cases}.\label{eq:equivalent-transform}
\end{equation}
We also define as $\mathbf{r}_{i}^{\prime}$ and $\mathbf{r}_{i}^{\prime\prime}$
the input signals (\ref{eq:received-signal-aggregated}) to RU $i$
under the assumption that the DPR scheme adopts the variables $\{\mathbf{L}_{e}^{\prime},\mathbf{\Omega}_{e}^{\prime}\}_{e\in\mathcal{E}_{\mathrm{act}}}$
and $\{\mathbf{I},\mathbf{\Omega}_{e}^{\prime\prime}\}_{e\in\mathcal{E}_{\mathrm{act}}}$,
respectively. We first prove a key relation between the signals $\mathbf{r}_{i}^{\prime}$
and $\mathbf{r}_{i}^{\prime\prime}$.

\begin{lemma}\label{lem:equality}Given (\ref{eq:Omega-double-prime})-(\ref{eq:equivalent-transform}),
the equalities

\begin{equation}
\mathbf{r}_{i}^{\prime\prime}=\begin{cases}
\mathrm{diag}\left(\{\mathbf{G}_{\tilde{e}}\}_{\tilde{e}\in\Gamma_{I}(i)}\right)\mathbf{r}_{i}^{\prime}, & \mathrm{if}\,\,\Gamma_{i}(i)\neq\varnothing\\
\mathbf{r}_{i}^{\prime}, & \mathrm{otherwise}
\end{cases}\label{eq:equality-ri}
\end{equation}
hold for all $i\in\mathcal{V}$.\end{lemma}\textit{Proof:} We prove
(\ref{eq:equality-ri}) by induction. It is straightforward to see
that (\ref{eq:equality-ri}) is true for any RUs $i$ in layers $\mathcal{V}_{1}$
and $\mathcal{V}_{2}$. The proof is completed by showing that if
the equalities $\mathbf{r}_{\mathrm{tail}(e)}^{\prime\prime}=\mathrm{diag}(\{\mathbf{G}_{\tilde{e}}\}_{\tilde{e}\in\Gamma_{I}(\mathrm{tail}(e))})\mathbf{r}_{\mathrm{tail}(e)}^{\prime}$
hold for all incoming edges $e\in\Gamma_{I}(i)$ of node $i$, then
we also have the equality
\begin{equation}
\mathbf{r}_{i}^{\prime\prime}=\mathrm{diag}\left(\{\mathbf{G}_{\tilde{e}}\}_{\tilde{e}\in\Gamma_{I}(i)}\right)\mathbf{r}_{i}^{\prime},\label{eq:equality-induction}
\end{equation}
for the next node $i$. The left-hand side of (\ref{eq:equality-induction})
is calculated as
\begin{align}
\mathbf{r}_{i}^{\prime\prime} & =\left[\begin{array}{c}
\mathbf{u}_{e_{1}^{i}}^{\prime\prime}\\
\vdots\\
\mathbf{u}_{e_{|\Gamma_{I}(i)|}^{i}}^{\prime\prime}
\end{array}\right]=\left[\begin{array}{c}
\mathbf{r}_{\mathrm{tail}(e_{1}^{i})}^{\prime\prime}\\
\vdots\\
\mathbf{r}_{\mathrm{tail}(e_{|\Gamma_{I}(i)|}^{i})}^{\prime\prime}
\end{array}\right]+\left[\begin{array}{c}
\mathbf{q}_{e_{1}^{i}}^{\prime\prime}\\
\vdots\\
\mathbf{q}_{e_{|\Gamma_{I}(i)|}^{i}}^{\prime\prime}
\end{array}\right]\\
 & =\left[\begin{array}{c}
\mathrm{diag}\left(\{\mathbf{G}_{\tilde{e}}\}_{\tilde{e}\in\Gamma_{I}(\mathrm{tail}(e_{1}^{i}))}\right)\mathbf{r}_{\mathrm{tail}(e_{1}^{i})}^{\prime}\\
\vdots\\
\mathrm{diag}\left(\{\mathbf{G}_{\tilde{e}}\}_{\tilde{e}\in\Gamma_{I}(\mathrm{tail}(e_{|\Gamma_{I}(i)|}^{i}))}\right)\mathbf{r}_{\mathrm{tail}(e_{|\Gamma_{I}(i)|}^{i})}^{\prime}
\end{array}\right]+\left[\begin{array}{c}
\mathbf{G}_{e_{1}^{i}}\mathbf{q}_{e_{1}^{i}}^{\prime}\\
\vdots\\
\mathbf{G}_{e_{|\Gamma_{I}(i)|}^{i}}\mathbf{q}_{e_{|\Gamma_{I}(i)|}^{i}}^{\prime}
\end{array}\right],\label{eq:LHS}
\end{align}
where $\mathbf{q}_{e}^{\prime}$ and $\mathbf{q}_{e}^{\prime\prime}$
are quantization noise signals obtained with the variables $\{\mathbf{L}_{e}^{\prime},\mathbf{\Omega}_{e}^{\prime}\}_{e\in\mathcal{E}_{\mathrm{act}}}$
and $\{\mathbf{I},\mathbf{\Omega}_{e}^{\prime\prime}\}_{e\in\mathcal{E}_{\mathrm{act}}}$,
respectively, and we recall the notation $\Gamma_{I}(i)=\{e_{1}^{i},\ldots,e_{|\Gamma_{I}(i)|}^{i}\}$.
Also, the right-hand side of (\ref{eq:equality-induction}) is given
as
\begin{align}
\mathrm{diag}\left(\{\mathbf{G}_{\tilde{e}}\}_{\tilde{e}\in\Gamma_{I}(i)}\right)\mathbf{r}_{i}^{\prime}= & \left[\begin{array}{c}
\mathbf{G}_{e_{1}^{i}}\mathbf{u}_{e_{1}^{i}}^{\prime}\\
\vdots\\
\mathbf{G}_{e_{|\Gamma_{I}(i)|}^{i}}\mathbf{u}_{e_{|\Gamma_{I}(i)|}^{i}}^{\prime}
\end{array}\right]\\
= & \left[\begin{array}{c}
\mathbf{G}_{e_{1}^{i}}\mathbf{L}_{e_{1}^{i}}^{\prime}\mathbf{r}_{\mathrm{tail}(e_{1}^{i})}^{\prime}\\
\vdots\\
\mathbf{G}_{e_{|\Gamma_{I}(i)|}^{i}}\mathbf{L}_{e_{|\Gamma_{I}(i)|}^{i}}^{\prime}\mathbf{r}_{\mathrm{tail}(e_{|\Gamma_{I}(i)|}^{i})}^{\prime}
\end{array}\right]+\left[\begin{array}{c}
\mathbf{G}_{e_{1}^{i}}\mathbf{q}_{e_{1}^{i}}^{\prime}\\
\vdots\\
\mathbf{G}_{e_{|\Gamma_{I}(i)|}^{i}}\mathbf{q}_{e_{|\Gamma_{I}(i)|}^{i}}^{\prime}
\end{array}\right]\\
= & \left[\begin{array}{c}
\mathrm{diag}\left(\{\mathbf{G}_{\tilde{e}}\}_{\tilde{e}\in\Gamma_{I}(\mathrm{tail}(e_{1}^{i}))}\right)(\mathbf{L}_{e_{1}^{i}}^{\prime})^{-1}\mathbf{L}_{e_{1}^{i}}^{\prime}\mathbf{r}_{\mathrm{tail}(e_{1}^{i})}^{\prime}\\
\vdots\\
\mathrm{diag}\left(\{\mathbf{G}_{\tilde{e}}\}_{\tilde{e}\in\Gamma_{I}(\mathrm{tail}(e_{|\Gamma_{I}(i)|}^{i}))}\right)(\mathbf{L}_{e_{|\Gamma_{I}(i)|}^{i}}^{\prime})^{-1}\mathbf{L}_{e_{|\Gamma_{I}(i)|}^{i}}^{\prime}\mathbf{r}_{\mathrm{tail}(e_{|\Gamma_{I}(i)|}^{i})}^{\prime}
\end{array}\right]\nonumber \\
 & +\left[\begin{array}{c}
\mathbf{G}_{e_{1}^{i}}\mathbf{q}_{e_{1}^{i}}^{\prime}\\
\vdots\\
\mathbf{G}_{e_{|\Gamma_{I}(i)|}^{i}}\mathbf{q}_{e_{|\Gamma_{I}(i)|}^{i}}^{\prime}
\end{array}\right]
\end{align}
which equals (\ref{eq:LHS}). Thus, we have proved that (\ref{eq:equality-induction})
is true and that the equality (\ref{eq:equality-ri}) holds.~~~~~~~$\square$

Using Lemma \ref{lem:equality}, we now prove the equality of the
backhaul rates
\begin{equation}
g_{e}^{\mathrm{DPR}}(\{\mathbf{L}_{e}^{\prime},\mathbf{\Omega}_{e}^{\prime}\}_{e\in\mathcal{E}_{\mathrm{act}}})=g_{e}^{\mathrm{DPR}}(\{\mathbf{I},\mathbf{\Omega}_{e}^{\prime\prime}\}_{e\in\mathcal{E}_{\mathrm{act}}})\label{eq:equality-backhaul}
\end{equation}
for all $e\in\Gamma_{O}(i)$ and $i\in\mathcal{N}_{R}$, which implies
that the matrices $\{\mathbf{I},\mathbf{\Omega}_{e}^{\prime\prime}\}_{e\in\mathcal{E}_{\mathrm{act}}}$
are feasible if matrices $\{\mathbf{L}_{e}^{\prime},\mathbf{\Omega}_{e}^{\prime}\}_{e\in\mathcal{E}_{\mathrm{act}}}$
are; and we also prove the equality of the sum-rates, i.e.,
\begin{equation}
f^{\mathrm{DPR}}(\{\mathbf{L}_{e}^{\prime},\mathbf{\Omega}_{e}^{\prime}\}_{e\in\mathcal{E}_{\mathrm{act}}})=f^{\mathrm{DPR}}(\{\mathbf{I},\mathbf{\Omega}_{e}^{\prime\prime}\}_{e\in\mathcal{E}_{\mathrm{act}}}).\label{eq:equality-sum-rates}
\end{equation}
These equalities will prove the claim in Proposition \ref{prop:identity}
for full-rank matrices $\mathbf{L}_{e}^{\prime}$.

To show (\ref{eq:equality-backhaul}), define as $\mathbf{u}_{e}^{\prime}$
and $\mathbf{u}_{e}^{\prime\prime}$ the compressed baseband signals
transmitted on edge $e$ with the variables $\{\mathbf{L}_{e}^{\prime},\mathbf{\Omega}_{e}^{\prime}\}_{e\in\mathcal{E}_{\mathrm{act}}}$
and $\{\mathbf{I},\mathbf{\Omega}_{e}^{\prime\prime}\}_{e\in\mathcal{E}_{\mathrm{act}}}$,
respectively. Then, by direct calculation using (\ref{eq:backhaul-constraint-DR}),
we get
\begin{align}
g_{e}^{\mathrm{DPR}}(\{\mathbf{L}_{e}^{\prime},\mathbf{\Omega}_{e}^{\prime}\}_{e\in\mathcal{E}_{\mathrm{act}}})= & \log\det\left(\mathbf{\Omega}_{e}^{\prime}+\mathbf{L}_{e}^{\prime}\mathbf{\Sigma}_{\mathbf{r}_{i}^{\prime}}(\mathbf{L}_{e}^{\prime})^{\dagger}\right)-\log\det\left(\mathbf{\Omega}_{e}^{\prime}\right),\\
\mathrm{and}\,\, g_{e}^{\mathrm{DPR}}(\{\mathbf{I},\mathbf{\Omega}_{e}^{\prime\prime}\}_{e\in\mathcal{E}_{\mathrm{act}}})= & \log\det\left(\mathbf{\Omega}_{e}^{\prime\prime}+\mathbf{\Sigma}_{\mathbf{r}_{i}^{\prime\prime}}\right)-\log\det\left(\mathbf{\Omega}_{e}^{\prime\prime}\right)\\
= & \log\det\left(\mathbf{G}_{e}\mathbf{\Omega}_{e}^{\prime}\mathbf{G}_{e}^{\dagger}+\mathrm{diag}\left(\{\mathbf{G}_{\tilde{e}}\}_{\tilde{e}\in\Gamma_{I}(i)}\right)\mathbf{\Sigma}_{\mathbf{r}_{i}^{\prime}}\mathrm{diag}\left(\{\mathbf{G}_{\tilde{e}}^{\dagger}\}_{\tilde{e}\in\Gamma_{I}(i)}\right)\right)\nonumber \\
 & -\log\det\left(\mathbf{G}_{e}\mathbf{\Omega}_{e}^{\prime}\mathbf{G}_{e}^{\dagger}\right)\\
= & \log\det\left((\mathbf{L}_{e}^{\prime})^{-1}\mathbf{\Omega}_{e}^{\prime}(\mathbf{L}_{e}^{\prime})^{-\dagger}+\mathbf{\Sigma}_{\mathbf{r}_{i}^{\prime}}\right)-\log\det\left((\mathbf{L}_{e}^{\prime})^{-1}\mathbf{\Omega}_{e}^{\prime}(\mathbf{L}_{e}^{\prime})^{-\dagger}\right)\\
= & \log\det\left(\mathbf{\Omega}_{e}^{\prime}+\mathbf{L}_{e}^{\prime}\mathbf{\Sigma}_{\mathbf{r}_{i}^{\prime}}(\mathbf{L}_{e}^{\prime})^{\dagger}\right)-\log\det\left(\mathbf{\Omega}_{e}^{\prime}\right).
\end{align}
Similarly, (\ref{eq:equality-sum-rates}) can be proved by direct
calculation.

While the proof provided above holds under the assumption that the
matrices $\mathbf{L}_{e}^{\prime}$ are full rank, Proposition \ref{prop:identity}
can be seen to hold more generally for rank-deficient matrices $\mathbf{L}_{e}^{\prime}$.
This follows by perturbing the matrices $\mathbf{L}_{e}^{\prime}$
in order to make them full rank (i.e., as $\mathbf{L}_{e}^{\prime}+\epsilon\mathbf{I}$),
and then using continuity of the functions $f^{\mathrm{DPR}}(\{\mathbf{L}_{e}^{\prime},\mathbf{\Omega}_{e}^{\prime}\}_{e\in\mathcal{E}_{\mathrm{act}}})$
and $g_{e}^{\mathrm{DPR}}(\{\mathbf{L}_{e}^{\prime},\mathbf{\Omega}_{e}^{\prime}\}_{e\in\mathcal{E}_{\mathrm{act}}})$
with respect to the variables $\mathbf{L}_{e}^{\prime}$.

\section{Calculation of the correlation matrices in (\ref{eq:conditional-covariance-1})-(\ref{eq:conditional-covariance-2})}\label{appendix:calculation-correlation}

In this appendix, we show how to compute the correlation matrices
$\mathbf{\Sigma}_{\mathbf{x},\mathbf{v}_{e}}$, $\mathbf{\Sigma}_{\tilde{\mathbf{n}}_{\pi_{l}(i)},\mathbf{v}_{e}}$
and $\mathbf{\Sigma}_{\mathbf{v}_{e}}$ appearing in (\ref{eq:conditional-covariance-1})-(\ref{eq:conditional-covariance-2}).
To this end, we first define as $\tilde{\mathcal{V}}_{\mathrm{head}(e)}=\{\tilde{v}_{1}^{\mathrm{head}(e)},\ldots,\tilde{v}_{|\mathcal{\tilde{V}}_{\mathrm{head}(e)}|}^{\mathrm{head}(e)}\}$
and $\tilde{\mathcal{E}}_{\mathrm{head}(e)}=\{\tilde{e}_{1}^{\mathrm{head}(e)},\ldots,\tilde{e}_{|\tilde{\mathcal{E}}_{\mathrm{head}(e)}|}^{\mathrm{head}(e)}\}$
the sets of the RUs and the edges belonging to the subnetwork consisting
of the RU $\mathrm{head}(e)$ and its ascendant nodes $\mathrm{ASC}(\mathrm{head}(e))$.
Then, the signals $\tilde{\mathbf{n}}_{\pi_{l}(i)}$ and $\mathbf{v}_{e}$
can be written as
\begin{align}
\tilde{\mathbf{n}}_{\pi_{l}(i)} & =\mathbf{T}_{\mathbf{r}_{\pi_{l}(i)}}^{Z}\mathbf{z}_{\tilde{\mathcal{V}}_{\mathrm{head}(e)}}+\mathbf{T}_{\mathbf{r}_{\pi_{l}(i)}}^{Q}\mathbf{q}_{\tilde{\mathcal{E}}_{\mathrm{head}(e)}},\\
\mathrm{and}\,\,\mathbf{v}_{e} & =\tilde{\mathbf{H}}_{\mathbf{v}_{e}}\mathbf{x}+\mathbf{T}_{\mathbf{v}_{e}}^{Z}\mathbf{z}_{\tilde{\mathcal{V}}_{\mathrm{head}(e)}}+\mathbf{T}_{\mathbf{v}_{e}}^{Q}\mathbf{q}_{\tilde{\mathcal{E}}_{\mathrm{head}(e)}},
\end{align}
where we have defined the matrices
\begin{align*}
\tilde{\mathbf{H}}_{\mathbf{v}_{e}}= & [\tilde{\mathbf{H}}_{e_{1}^{\mathcal{S}_{e}}};\ldots;\tilde{\mathbf{H}}_{e_{|\mathcal{S}_{e}|}^{\mathcal{S}_{e}}}],\\
\tilde{\mathbf{H}}_{e^{\prime}}= & \begin{cases}
\mathbf{H}_{\mathrm{tail}(e^{\prime})}, & \,\mathrm{if}\,\,\Gamma_{I}(\mathrm{tail}(e^{\prime}))=\textrm{�}\\
{}[\tilde{\mathbf{H}}_{e_{1}^{\mathrm{tail}(e^{\prime})}};\ldots;\tilde{\mathbf{H}}_{e_{|\Gamma_{I}(\mathrm{tail}(e^{\prime}))|}^{\mathrm{tail}(e^{\prime})}}], & \,\mathrm{otherwise}
\end{cases},\\
\mathbf{T}_{\mathbf{r}_{\pi_{l}(i)}}^{Z}= & [(\mathbf{E}_{\pi_{l}(i)}^{Z})^{\dagger};\mathbf{T}_{e_{1}^{\pi_{l}(i)}}^{Z};\ldots;\mathbf{T}_{e_{|\Gamma_{I}(\pi_{l}(i))|}^{\pi_{l}(i)}}^{Z}],\\
\mathbf{T}_{\mathbf{r}_{\pi_{l}(i)}}^{Q}= & [\mathbf{T}_{e_{1}^{\pi_{l}(i)}}^{Q};\ldots;\mathbf{T}_{e_{|\Gamma_{I}(\pi_{l}(i))|}^{\pi_{l}(i)}}^{Q}],\\
\mathbf{T}_{\mathbf{v}_{e}}^{Z}= & [\mathbf{T}_{e_{1}^{\mathcal{S}_{e}}}^{Z};\ldots;\mathbf{T}_{e_{|\mathcal{S}_{e}|}^{\mathcal{S}_{e}}}^{Z}],\\
\mathrm{and}\,\,\mathbf{T}_{\mathbf{v}_{e}}^{Q}= & [\mathbf{T}_{e_{1}^{\mathcal{S}_{e}}}^{Q};\ldots;\mathbf{T}_{e_{|\mathcal{S}_{e}|}^{\mathcal{S}_{e}}}^{Q}],
\end{align*}
with the notation $\mathcal{S}_{e}=\{e_{1}^{\mathcal{S}_{e}},\ldots,e_{|\mathcal{S}_{e}|}^{\mathcal{S}_{e}}\}$
and the matrices
\begin{align*}
\mathbf{T}_{e^{\prime}}^{Z} & =\begin{cases}
(\mathbf{E}_{\mathrm{tail}(e^{\prime})}^{Z})^{\dagger}, & \mathrm{if}\,\,\Gamma_{I}(\mathrm{tail}(e^{\prime}))=\textrm{�}\\
{}[(\mathbf{E}_{\mathrm{tail}(e^{\prime})}^{Z})^{\dagger};\mathbf{T}_{e_{1}^{\mathrm{tail}(e^{\prime})}}^{Z};\ldots;\mathbf{T}_{e_{|\Gamma_{I}(\mathrm{tail}(e^{\prime}))|}^{\mathrm{tail}(e^{\prime})}}^{Z}], & \mathrm{otherwise}
\end{cases},\\
\mathrm{and}\,\,\mathbf{T}_{e^{\prime}}^{Q} & =\begin{cases}
(\mathbf{E}_{e^{\prime}}^{Q})^{\dagger}, & \mathrm{if}\,\,\Gamma_{I}(\mathrm{tail}(e^{\prime}))=\textrm{�}\\
{}[(\mathbf{E}_{e^{\prime}}^{Q})^{\dagger};\mathbf{T}_{e_{1}^{\mathrm{tail}(e^{\prime})}}^{Q};\ldots;\mathbf{T}_{e_{|\Gamma_{I}(\mathrm{tail}(e^{\prime}))|}^{\mathrm{tail}(e^{\prime})}}^{Q}], & \mathrm{otherwise}
\end{cases}.
\end{align*}
Here, we have defined the matrix $\mathbf{E}_{\tilde{v}_{m}^{\mathrm{head}(e)}}^{Z}\in\mathbb{C}^{(\sum_{j\in\tilde{\mathcal{V}}_{\mathrm{head}(e)}}n_{R,j})\times n_{R,\tilde{v}_{m}^{\mathrm{head}(e)}}}$
having all zero elements except for the rows from $(\sum_{j=1}^{m-1}n_{R,\tilde{v}_{j}^{\mathrm{head}(e)}}+1)$
to $(\sum_{j=1}^{m}n_{R,\tilde{v}_{j}^{\mathrm{head}(e)}})$ which
contain an $n_{R,\tilde{v}_{m}^{\mathrm{head}(e)}}\times n_{R,\tilde{v}_{m}^{\mathrm{head}(e)}}$
identity matrix, and the matrix $\mathbf{E}_{\tilde{e}_{l}^{\mathrm{head}(e)}}^{Q}\in\mathbb{C}^{(\sum_{e^{\prime}\in\tilde{\mathcal{E}}_{\mathrm{head}(e)}}d_{e^{\prime}})\times d_{\tilde{e}_{l}^{\mathrm{head}(e)}}}$
having all zero elements except for the rows from $(\sum_{j=1}^{l-1}d_{\tilde{e}_{j}^{\mathrm{head}(e)}}+1)$
to $(\sum_{j=1}^{l}d_{\tilde{e}_{j}^{\mathrm{head}(e)}})$ which contain
an $d_{\tilde{e}_{l}^{\mathrm{head}(e)}}\times d_{\tilde{e}_{l}^{\mathrm{head}(e)}}$
identity matrix.

As a result, the correlation matrices $\mathbf{\Sigma}_{\mathbf{x},\mathbf{v}_{e}}$,
$\mathbf{\Sigma}_{\tilde{\mathbf{n}}_{\pi_{l}(i)},\mathbf{v}_{e}}$
and $\mathbf{\Sigma}_{\mathbf{v}_{e}}$ can be computed as
\begin{align}
 & \mathbf{\Sigma}_{\mathbf{x},\mathbf{v}_{e}}=\mathbf{\Sigma}_{\mathbf{x}}\tilde{\mathbf{H}}_{\mathbf{v}_{e}}^{\dagger},\\
 & \mathbf{\Sigma}_{\tilde{\mathbf{n}}_{\pi_{l}(i)},\mathbf{v}_{e}}=\mathbf{T}_{\mathbf{r}_{\pi_{l}(i)}}^{Z}(\mathbf{T}_{\mathbf{v}_{e}}^{Z})^{\dagger}+\mathbf{T}_{\mathbf{r}_{\pi_{l}(i)}}^{Q}\mathrm{diag}(\{\mathbf{\Omega}_{e^{\prime}}\}_{e^{\prime}\in\tilde{\mathcal{E}}_{\mathrm{head}(e)}})(\mathbf{T}_{\mathbf{v}_{e}}^{Q})^{\dagger},\\
\mathrm{and}\,\, & \mathbf{\Sigma}_{\mathbf{v}_{e}}=\tilde{\mathbf{H}}_{\mathbf{v}_{e}}\mathbf{\Sigma}_{\mathbf{x}}\tilde{\mathbf{H}}_{\mathbf{v}_{e}}^{\dagger}+\mathbf{T}_{\mathbf{v}_{e}}^{Z}(\mathbf{T}_{\mathbf{v}_{e}}^{Z})^{\dagger}+\mathbf{T}_{\mathbf{v}_{e}}^{Q}\mathrm{diag}(\{\mathbf{\Omega}_{e^{\prime}}\}_{e^{\prime}\in\tilde{\mathcal{E}}_{\mathrm{head}(e)}})(\mathbf{T}_{\mathbf{v}_{e}}^{Q})^{\dagger}.
\end{align}

\end{document}